\documentclass[a4paper,12pt]{article}
\pdfoutput=1
\usepackage{epsfig}
\usepackage{amssymb}
\usepackage{amsfonts}
\usepackage{amsmath}
\usepackage{euscript}
\usepackage{verbatim}
\usepackage{soul}
\usepackage[normalem]{ulem}
\usepackage{latexsym}
\usepackage{graphicx}
\usepackage{caption}
\usepackage{float}
\usepackage{subcaption}
\usepackage[hidelinks]{hyperref}
\hypersetup{
    colorlinks=true,
    linkcolor=black,
    citecolor=black,
    urlcolor=black,
    pdfborder={0 0 0},     
    pdfhighlight=/N         
}

\usepackage{amsmath,amssymb,booktabs,geometry,longtable}

\usepackage{listings}

\usepackage{booktabs}
\usepackage{appendix}

\newif\ifdtup

\catcode`\@=11

\@addtoreset{equation}{section}

\def\@normalsize{\@setsize\normalsize{15pt}\xiipt\@xiipt
\abovedisplayskip 14pt plus3pt minus3pt%
\belowdisplayskip \abovedisplayskip
\abovedisplayshortskip \z@ plus3pt%
\belowdisplayshortskip 7pt plus3.5pt minus0pt}

\def\small{\@setsize\small{13.6pt}\xipt\@xipt
\abovedisplayskip 13pt plus3pt minus3pt%
\belowdisplayskip \abovedisplayskip
\abovedisplayshortskip \z@ plus3pt%
\belowdisplayshortskip 7pt plus3.5pt minus0pt
\def\@listi{\parsep 4.5pt plus 2pt minus 1pt
     \itemsep \parsep
     \topsep 9pt plus 3pt minus 3pt}}

\relax

\catcode`@=12

\catcode`\@=11

\def\section{\@startsection{section}{1}{\z@}{3.5ex plus 1ex minus
   .2ex}{2.3ex plus .2ex}{\large\bf}}

\def\SymBoxes#1#2#3#4{\newdimen\un@t \un@t#3%
\raisebox{#1}{\rule{#2\un@t}{#4}\hskip-#2\un@t
\@tempdimb\un@t \advance\@tempdimb by-#4\@tempcntb#2\relax%
\@whilenum{\@tempcntb>0}\do{
\rule{#4}{\un@t}\hskip\@tempdimb \advance\@tempcntb by\m@ne}%
\hskip-#2\un@t \rule[\un@t]{#2\un@t}{#4}%
\rule[\un@t]{#4}{#4}\hskip-#4
\rule{#4}{\un@t}}\hskip-#4}                

\begin{document}

\newcommand{\beq}{\begin{equation}}
\newcommand{\eeq}{\end{equation}}
\newcommand{\bea}{\begin{eqnarray}}
\newcommand{\eea}{\end{eqnarray}}
\newcommand{\beas}{\begin{eqnarray*}}
\newcommand{\eeas}{\end{eqnarray*}}
\newcommand{\defi}{\stackrel{\rm def}{=}}
\newcommand{\non}{\nonumber}
\newcommand{\bquo}{\begin{quote}}
\newcommand{\enqu}{\end{quote}}
\renewcommand{\(}{\beq}
\renewcommand{\)}{\eeq}
\def \eqn#1#2{\beq#2\label{#1}\eeq}

\def\e{\epsilon}
\def\IZ{{\mathbb Z}}
\def\IR{{\mathbb R}}
\def\IC{{\mathbb C}}
\def\IQ{{\mathbb Q}}
\def\de{\partial}
\def\Tr{ \hbox{\rm Tr}}
\def\H{ \hbox{\rm H}}
\def\HE{ \hbox{$\rm H^{even}$}}
\def\HO{ \hbox{$\rm H^{odd}$}}
\def\K{ \hbox{\rm K}}
\def\Im{ \hbox{\rm Im}}
\def\Ker{ \hbox{\rm Ker}}
\def\const{\hbox {\rm const.}}
\def\o{\over}
\def\im{\hbox{\rm Im}}
\def\re{\hbox{\rm Re}}
\def\bra{\langle}\def\ket{\rangle}
\def\Arg{\hbox {\rm Arg}}
\def\Re{\hbox {\rm Re}}
\def\Im{\hbox {\rm Im}}
\def\exo{\hbox {\rm exp}}
\def\diag{\hbox{\rm diag}}
\def\longvert{{\rule[-2mm]{0.1mm}{7mm}}\,}
\def\a{\alpha}
\def\dag{{}^{\dagger}}
\def\tq{{\widetilde q}}
\def\p{{}^{\prime}}
\def\W{W}
\def\N{{\cal N}}
\def\hsp{,\hspace{.7cm}}

\def\br{\nonumber}
\def\IZ{{\mathbb Z}}
\def\IR{{\mathbb R}}
\def\IC{{\mathbb C}}
\def\IQ{{\mathbb Q}}
\def\IP{{\mathbb P}}
\def \eqn#1#2{\beq#2\label{#1}\eeq}

\newcommand{\C}{\ensuremath{\mathbb C}}
\newcommand{\Z}{\ensuremath{\mathbb Z}}
\newcommand{\R}{\ensuremath{\mathbb R}}
\newcommand{\rp}{\ensuremath{\mathbb {RP}}}
\newcommand{\cp}{\ensuremath{\mathbb {CP}}}
\newcommand{\vac}{\ensuremath{|0\rangle}}
\newcommand{\vact}{\ensuremath{|00\rangle}                    }
\newcommand{\oc}{\ensuremath{\overline{c}}}
\newcommand{\psizero}{\psi_{0}}
\newcommand{\phizero}{\phi_{0}}
\newcommand{\hzero}{h_{0}}
\newcommand{\psiin}{\psi_{\rh}}
\newcommand{\phiin}{\phi_{\rh}}
\newcommand{\hin}{h_{\rh}}
\newcommand{\rh}{r_{h}}
\newcommand{\rb}{r_{b}}
\newcommand{\psibnd}{\psi_{0}^{b}}
\newcommand{\psibndp}{\psi_{1}^{b}}
\newcommand{\phibnd}{\phi_{0}^{b}}
\newcommand{\phibndp}{\phi_{1}^{b}}
\newcommand{\gbnd}{g_{0}^{b}}
\newcommand{\hbnd}{h_{0}^{b}}
\newcommand{\zh}{z_{h}}
\newcommand{\zb}{z_{b}}
\newcommand{\man}{\mathcal{M}}
\newcommand{\hbr}{\bar{h}}
\newcommand{\tbr}{\bar{t}}

\begin{titlepage}

\def\thefootnote{\fnsymbol{footnote}}

\begin{center}
{\large
{\bf Wilson Towers as Local Bulk Fields
}}
\end{center}

\begin{center}
\hspace{0.2cm} 
Vishal Gayari$^a$\footnote{\texttt{vishalgayari@iisc.ac.in}},  Chethan Krishnan$^a$\footnote{\texttt{chethan.krishnan.physics@gmail.com}}
\end{center}

\renewcommand{\thefootnote}{\arabic{footnote}}

\begin{center}

$^a$ {Center for High Energy Physics,\\
Indian Institute of Science, Bangalore 560012, India}\\

\end{center}
\vspace{-0.15in}
\noindent

\begin{center} {\bf Abstract} \end{center}
Multi-winding Wilson loops (Wilson spools) in 2+1-dimensional gravity can reproduce one-loop partition functions of local free fields in the bulk. Bulk free fields have a second-quantized Fock space, and the AdS/CFT correspondence suggests that the associated multi-particle sectors are dual to multi-trace primaries in the CFT. In this note, we use standard symmetric-function methods to show that the Wilson spool in thermal AdS$_3$ can be recast as a sum over {\em single}-winding Wilson loops -- one for each multi-trace primary. In a companion paper to appear, we will view Wilson networks and TQFTs as the natural language of non-perturbative bulk quantum gravity. The present note illustrates how this can apply to {\em local} bulk fields, and not just defects: a bulk (generalized free) field is to be viewed as a full tower of multi-trace Wilson lines. We further show that the $SL(2)$ descendants of each multi-trace primary, together with the boundary gravitons of the AdS$_3$ background, correctly reproduce the full Virasoro character of each module. In this language, the role of a light insertion on a heavy primary is played by a topological Verlinde line. This allows us to obtain a ``microscopic" description of one-loop determinants on the smooth BTZ handlebody. A spatially wound probe Wilson line on a torus with a contractible {\em thermal} cycle can be traded for a Verlinde line inserted on a dense family of heavy Polyakov loops -- with the roles of the two cycles exchanged, so that the {\em spatial} cycle is now contractible. The vacuum row of the modular $S$-kernel acts as the (approximate) density of the heavy primaries. This reinforces the case made in arXiv:2601.18775 that a smooth horizon is a stand-in for an ensemble of quantum states, each produced by a heavy Wilson line that appears as a singular horizon in the semi-classical limit. 



\vspace{1.6 cm}
\vfill

\end{titlepage}

\setcounter{footnote}{0}

\tableofcontents

\section{Introduction and Summary}
\label{sec:intro}

Asymptotically AdS$_3$ quantum gravity admits a description built on an underlying topological
skeleton \cite{KashaevQT,PonsotTeschner1,PonsotTeschner2,Teschner,AndersenKashaev,CEZ1,CEZ2,Pradipta,Vishal}. In this language the bulk is
assembled from Wilson networks supported on handlebody fillings of the boundary surface. A
network evaluated in a definite choice of representations computes a finite-$c$ Virasoro conformal
block, and the data that distinguish one holographic CFT from another --- its spectrum of Virasoro
primaries $\sigma$ and its OPE coefficients $C$ --- enter only as the labels carried by the Wilson
lines and the weights assigned to their junctions. Crossing symmetry and modular invariance are understood as the statement that the network sum is independent of the channel in which the bulk is decomposed. In a companion paper \cite{Vishal}, we argue that this Wilson-network/TQFT skeleton language provides a useful non-perturbative formulation of {\it individual} holographic CFTs\footnote{Note that even though the ingredients are closely related, this perspective is different from that adopted in the Virasoro TQFT program \cite{CEZ1,CEZ2}: there the goal is to understand pure 3D gravity.}. Heavy
microstates and defects are accommodated in this picture as Wilson lines in appropriate representations \cite{Pradipta,Vishal,Rajdeep,DeserJackiwtHooft,Witten1988,Matschull, AmmonCastroIqbal,dBJ,CastroIqbalLlabres}.
It is less immediate how the {\it local}, propagating bulk fields of the low-energy effective
description are to be represented in the same Wilson-line language. The purpose of this note is to
make this explicit for free bulk fields: we will see that their one-loop determinants fit into the
skeleton in a way that is revealing of the general picture.

The torus is the simplest setting in which to pose the question. The partition function of a 2D CFT
at modular parameter $\tau$ decomposes into Virasoro characters,
\begin{equation}
Z(\tau,\bar\tau)
\;=\;
\sum_{h,\bar h} N_{h,\bar h}\,\chi_h(\tau)\,\bar\chi_{\bar h}(\bar\tau),
\qquad q\defi e^{2\pi i\tau},
\label{eq:intro_Zchar}
\end{equation}
with non-negative integer multiplicities $N_{h,\bar h}$ that encode the (discrete) spectrum. In the
TQFT skeleton each (paired non-chiral) character in \eqref{eq:intro_Zchar} is computed by a {\it single}-winding Wilson loop  in the representation labeled by the corresponding primary,
wrapping the non-contractible cycle of the solid torus. The
partition function is therefore a sum of single-winding Wilson loops, one per primary.

We are interested in holographic CFTs, for which the relevant regime is large central charge $c$
(equivalently large $N$) together with a sparse spectrum of light operators. Among the light
operators are the single-trace primaries $\mathcal{O}_i$ corresponding to the propagating bulk fields. To
leading order in $1/c$ each such operator is a {\it generalized free field}: its connected
higher-point functions are suppressed, and the operators built from it organize into the Fock space
of a free field \cite{Greenberg,HPPS,FKPS,ElShowkPapadodimas}. Consistency of the OPE then requires the spectrum to contain,
alongside each single-trace primary, an entire tower of light {\it multi-trace} primaries
$:\!\mathcal{O}_{i_1}\!\cdots\mathcal{O}_{i_k}\!:$ (suitably dressed with derivatives) together with
their descendants.\footnote{We use ``single-trace'' and ``multi-trace'' in their standard AdS/CFT
sense, and not in reference to any particular large-$N$ gauge theory. In the gauge-theory setting a
single-trace operator $\mathrm{Tr}(\cdots)$ is dual to a single-particle bulk state and a
normal-ordered product of traces to a multi-particle state \cite{AharonyBerkoozSilverstein,WittenMultiTrace}. Without committing to an explicit gauge theory,
``single-trace'' should be understood as the seed primary, and ``multi-trace'' denotes the composite primaries built as normal-ordered products of that field and its derivatives, i.e.\ the multi-particle states of the bulk Fock space. We will not carry around explicit gauge traces.} The multiplicities with which these multi-trace primaries appear in \eqref{eq:intro_Zchar}
are fixed entirely by the single-particle data.

It is useful to make the tower explicit for the  illustrative case of a single bulk scalar, dual to a single-trace primary
$\mathcal{O}$ of weight $(h,h)$. The one-particle states are the global $SL(2)_L\times SL(2)_R$
descendants of $\mathcal{O}$, spanning the module $\mathcal{V}_h\otimes\bar{\mathcal{V}}_h$ with
character
\begin{equation}
\chi_1(q,\bar q)\;=\;\frac{q^h\,\bar q^h}{(1-q)(1-\bar q)}.
\label{eq:intro_seed}
\end{equation}
The multi-particle states --- precisely the multi-trace primaries and their global descendants --- are the
symmetric powers of this module, and their generating function is the bosonic Fock-space character obtained via plethystic exponentiation \cite{Sundborg,AMMPV}:
\begin{equation}
Z_{\mathcal{O}}(\tau,\bar\tau)\;=\;
\exp\!\Big(\sum_{n=1}^{\infty}\frac{1}{n}\,\chi_1(q^n,\bar q^n)\Big)
\;=\;
\prod_{\ell,\bar\ell=0}^{\infty}\frac{1}{1-q^{\ell+h}\,\bar q^{\bar\ell+h}}.
\label{eq:intro_PE}
\end{equation}

The second equality in \eqref{eq:intro_PE} is a mathematical identity. It can be viewed as demonstrating that this same object is also the bulk one loop determinant of the dual scalar field on thermal AdS$_3$ \cite{GMY}. This is a manifestation of the AdS/CFT philosophy that we mentioned earlier: weakly coupled bulk fields are dual to towers of multi-trace primaries in the CFT. The {\it total} contribution of the field $\mathcal{O}$ to \eqref{eq:intro_Zchar} ---
the sum of the character of $\mathcal{O}$ and those of all its multi-trace progeny --- should be the bulk field's one-loop determinant. Read through the TQFT skeleton perspective, in which each primary character is a single Wilson loop, this says that a single {\it local} bulk field is not one Wilson loop but a {\it tower} of single-winding Wilson loops, one for each multi-trace primary it generates. We will call this the {\em Wilson tower}.

The first half of this note turns this expectation into a concrete rewriting, using the known
equivalence between the bulk one-loop determinant and a {\it Wilson spool} \cite{WilsonSpool,
BourneFlissKnighton}. The Wilson spool presents $Z_{\mathcal{O}}$ as a sum over {\it multiple
windings} of a single Wilson loop in the lowest-weight representation of weight $h$: the exponent in
\eqref{eq:intro_PE} is precisely this sum, the $n$-th term being an $n$-fold winding of the loop
around the thermal cycle of $\mathbb{H}^3/\mathbb{Z}$. Using standard symmetric-function
(Frobenius--Schur--Weyl) methods we resum this multi-winding series into a sum of {\it single}-winding
Wilson loops in higher representations. At fixed particle number $N$,
\begin{equation}
\chi_N(q,\bar q)
\;=\;
\sum_{m,\bar m\ge0} d^{(N)}_{m,\bar m}\,W_{(Nh+m,\;Nh+\bar m)}(q,\bar q),
\qquad
W_{(h_P,\bar h_P)}(q,\bar q)=\frac{q^{h_P}\bar q^{\bar h_P}}{(1-q)(1-\bar q)},
\label{eq:intro_main}
\end{equation}
where each single-winding loop $W_{(h_P,\bar h_P)}$ evaluates to the global character of the
corresponding representation and $d^{(N)}_{m,\bar m}$ counts the independent multi-trace primaries at
that weight. Summing over $N$ reproduces \eqref{eq:intro_PE}. The Wilson spool is thus literally a re-writing of the
Wilson tower anticipated above: the local bulk field is traded for a collection of single Wilson
loops, one per multi-trace primary. We make the dictionary fully explicit at low particle number
$N=2,3$, where an intermediate Young-diagram basis makes the counting transparent.

The characters $W_{(h_P,\bar h_P)}$ appearing in \eqref{eq:intro_main} are {\it global}
$SL(2)_L\times SL(2)_R$ characters, whereas the entries of \eqref{eq:intro_Zchar} are full {\it
Virasoro} characters. The difference is supplied by the boundary gravitons of the AdS$_3$ background.
The global module already contains the level-one descendants generated by $L_{-1}$ and $\bar L_{-1}$.
The remaining Virasoro descendants, generated by $L_{-n}$ with $n\ge2$, are universal and combine
multiplicatively with each global character to promote it to the corresponding Virasoro character.
Each multi-trace primary in the tower therefore contributes a complete Virasoro module to the
partition function, as required if \eqref{eq:intro_main} is to be read as a piece of
\eqref{eq:intro_Zchar}.

We then extend the discussion from thermal AdS$_3$ to the Euclidean BTZ handlebody. The two
geometries are the two solid-torus fillings of the same boundary torus, related by the modular $S$
transformation $\tau\to-1/\tau$: thermal AdS$_3$ has a contractible spatial cycle, while BTZ has a
contractible thermal cycle.\footnote{Throughout, a cycle is ``contractible'' if it bounds a disc in
the chosen solid-torus filling. We retain this terminology when a Wilson line is inserted along the
core of the filling: the core line is then linked by the contractible meridian.} In the skeleton
language the smooth BTZ handlebody is not a single geometry but a coarse-grained sum over heavy
Virasoro primaries, each represented by a heavy (continuous-series) Wilson line \cite{Pradipta, Vishal, Rajdeep}\footnote{This picture applies in Lorentzian signature as well. A Wilson line in the principal continuous series along the core of a semi-infinite solid Euclidean cylinder, in $SL(2,\IR)_k \times SL(2,\IR)_k$ Chern-Simons theory, prepares a heavy Virasoro primary above the BTZ threshold, when the cylinder is cut open. An interesting technicality is that one has to do a Drinfel'd-Sokolov reduction on the affine module, to obtain a genuine Virasoro module \cite{Rajdeep}. This reduction has the satisfying interpretation as the usual asymptotically AdS$_3$ demand.} and weighted by the
universal density supplied by the vacuum row of the modular $S$-kernel, i.e.\ the Cardy density
\cite{Cardy,HartmanKellerStoica,KrausMaloney}. 

The Wilson line/TQFT perspective suggests that a single probe Wilson line on this background is
captured, on each heavy microstate, by a topological Verlinde line $\mathsf D_a$ linking the heavy
Wilson-line source. The associated object is the defect-twined character of the heavy module,
\begin{equation}
\mathcal{Z}^{\rm tw}_{a|P}(\tau)
\;=\;
\Tr_{\mathcal{V}_P}\!\big(\mathsf D_a\,q^{L_0-c/24}\big)
\;=\;
\frac{S_{aP}}{S_{0P}}\,\widehat\chi_P(\tau),
\label{eq:intro_twined}
\end{equation}
the eigenvalue $S_{aP}/S_{0P}$ being the monodromy of the light label $a$ around a heavy primary of
momentum $P$ \cite{CEZ1,Verlinde,PetkovaZuber,BuicanGromovDefects}. Summing over the heavy spectrum
against the Cardy density $S_{0P}$ collapses the two kernels,
\begin{equation}
\mathcal{Z}^{\rm tw}_a(\tau)
\;=\;
\int_0^\infty dP\,S_{0P}\,\frac{S_{aP}}{S_{0P}}\,\widehat\chi_P(\tau)
\;=\;
\int_0^\infty dP\,S_{aP}\,\widehat\chi_P(\tau)
\;=\;
\widehat\chi_a\!\Big(-\tfrac{1}{\tau}\Big),
\label{eq:intro_cycleexchange}
\end{equation}
which is the dual-channel character of the light primary (we display the chiral half, the full answer
includes the antiholomorphic factor). The light line, initially a meridian linking the heavy source,
is thereby exchanged into a loop wrapping the complementary cycle of the handlebody --- the spatially
wound probe expected on BTZ. A plethystic tower of such probes reproduces the one-loop determinant on
BTZ in the same form obtained on thermal AdS$_3$, so the same spool-versus-tower correspondence holds
here. The construction gives a ``microscopic'' reading of the one-loop determinant on the smooth BTZ
background: it is assembled from probe towers on each heavy microstate, summed over microstates. It
reinforces the picture of \cite{Pradipta} (see also \cite{Rajdeep,Mohan,Datar,Burman1,Burman2}) that a smooth horizon
is a stand-in for an ensemble of quantum states --- an ensemble of microstates within a single CFT,
and not an ensemble of theories --- each produced by a heavy Wilson line that appears as a singular
horizon in the semiclassical limit.

Before concluding this section, let us mention some points of perspective. The question of coupling light fields minimally to Einstein gravity in 2+1 dimensions is an interesting one, see a very recent work in \cite{Bourne}. The TQFT skeleton picture asks one instead to identify a sewing/bootstrap consistent spectrum. Once the spectrum and OPE coefficients are identified, the Wilson network/TQFT specifies the complete dynamics. Since computing explicit Virasoro blocks is a tall order, even after simply assuming that the primary is part of the spectrum\footnote{Note that this assumption is inherent in all EFT-first approaches to coupling fields to gravity.}, this is hardly a flawless victory. But it has the virtue that it offers conceptual clarity: it reformulates the question as, ``what is the TQFT translation of the principle of equivalence"? Since minimal coupling is a universal operation in the metric language, this is a natural question for the TQFT in terms of properties of Wilson networks. From this point of view, the universality is reasonable because it corresponds to the fact that the trivalent coupling of a primary $i$ is controlled by $C_{i i^{\vee}\mathcal{V}_0} $ where $\mathcal{V}_0$ includes contributions from vacuum descendants (stress tensor exchanges) as well. Because $C_{i i^{\vee}\bf{1}}=1$, and the Ward identities fix the rest of the couplings through the same scaling dimension and spin that characterize the primary's kinematics, there is no independent species-dependent gravitational charge. This can be viewed as the origin of the equivalence principle. It also operationalizes the distinction between matter and ``pure" gravity in 2+1 dimensions in terms of the spectrum: ordinary matter is not a descendant of the vacuum. 

The remainder of the paper is organized as follows. In Section~\ref{sec:wilson_towers} we treat the
single bulk scalar on thermal AdS$_3$: we recall the GMY determinant and its Wilson-spool form
(Section~\ref{subsec:gmy_fliss_review}), pass from the fixed-$N$ multi-particle characters to the
multi-trace primaries (Sections~\ref{subsec:fixedN_to_primaries}--\ref{subsec:N3_counting}), and
assemble the resulting single-winding Wilson loops into the Wilson tower
(Section~\ref{subsec:multitrace_narrative}). Section~\ref{subsec:virasoro_dressing} shows how the
boundary gravitons dress the global characters into Virasoro characters, which contribute directly to the CFT partition function. Section~\ref{sec:heavy_light_cycle_exchange} treats the BTZ handlebody: we motivate the
defect-twined character of a heavy primary (Section~\ref{subsec:defect_twined_character}) and sum over
the heavy spectrum to obtain the cycle-exchanged light character
(Section~\ref{subsec:heavy_sum_defect_twined}). Appendix~\ref{sec:Multi_trace_primary_construction} and \ref{sec:honest_primaries}
count and construct the multi-trace primaries explicitly for $N=2,3$, and
Appendix~\ref{sec:schur_weyl_review} presents the equivalent organization of the fixed-$N$
partition functions in terms of windings and Young diagrams via the cycle-index and Cauchy
identities.

\section{Wilson Towers}
\label{sec:wilson_towers}


In this section we make precise the structural relationship between three {\it a priori} different organizational schemes for the same object -- the one-loop scalar partition function on thermal $\mathrm{AdS}_3\simeq \mathbb{H}^3/\mathbb{Z}$. We will track how the partition function moves between three distinct frameworks:
\begin{enumerate}
\item {\bf Winding Basis:} the Wilson-spool/heat-kernel expression for $\log Z$ naturally organizes the answer as a sum of $n$-winding traces, i.e.\ traces of the higher thermal holonomies $U^n$.
\item {\bf Multi-Trace Primary Basis:} the Hilbert-space trace interpretation of $Z$ is naturally a sum over irreducible $SL(2)_L\times SL(2)_R$ modules, i.e.\ one contribution per (generalized-free) multi-trace primary together with its global descendants.
\item {\bf Young-Diagram Basis:} at fixed particle number $N$, the same data can be repackaged as a sum over irreducible $S_N$ channels labeled by Young diagrams $\lambda\vdash N$. This basis trades ``windings for $S_N$-representations'': multi-winding traces can be re-expressed in terms of characters of representations of $S_N$.
\end{enumerate}
Our primary goal in this section will be to provide a translation between the first and second frameworks. We will show that the tower of multi-winding Wilson loops of a single seed primary, has a one to one correspondence with the tower of single-winding Wilson loops in the ``multi-trace" representations\footnote{We  will explain the details of doing this, later in this section.} of $SL(2)_L\times SL(2)_R$ constructed from the seed. We will also discuss the closely related Young-diagram basis in Appendix \ref{sec:schur_weyl_review}. The key distinction between these two approaches is that the multi-trace primary basis uses the characters of reps of $SL(2)_L\times SL(2)_R$ to organize the multi-windings, while in the Young-diagram basis the multi-windings are organized in term of characters of the representations of $S_N$.

Our main goal in this section is mostly calculational: we want to show that the multiplicities appearing in the fixed-$N$ spectrum are {\it precisely} the number of independent multi-trace primaries at the corresponding weights. We will make an effort to be concrete in many aspects of the calculations, including in the construction of  explicit low-lying primaries in the Appendices. 

We work with a single bulk scalar and a single primitive cycle (the thermal one).  In Section \ref{sec:heavy_light_cycle_exchange} we discuss the analogous situation for the BTZ handlebody -- this is interesting, because it allows us to give a ``microscopic" understanding of why the one loop determinant organizes itself in terms of loops around the spatial cycle. 
The more general statements for multiple primitive cycles should be straightforward: our focus here is on the basic idea rather than generality.

\subsection{Bulk Determinants and Wilson Spools}
\label{subsec:gmy_fliss_review}

We begin by recalling the basic scalar result on thermal $\mathrm{AdS}_3\simeq \mathbb{H}^3/\mathbb{Z}$ \cite{GMY,WilsonSpool,BourneFlissKnighton}. Let the complex boundary modulus $\tau = \tau_1 + i \tau_2$ define the standard coordinates
\begin{equation}
q = e^{2\pi i\tau}, \qquad \bar q = e^{-2\pi i\bar\tau}.
\label{eq:q_def}
\end{equation}
Giombi, Maloney, and Yin (GMY) computed the scalar determinant on the quotient $\mathbb{H}^3/\mathbb{Z}$ using heat kernel methods and obtained the winding-series expression \cite{GMY}
\begin{equation}
-\log\det\Delta
=
2\sum_{n=1}^\infty \frac{|q|^{2nh}}{n\,|1-q^n|^2},
\qquad \text{where} \qquad
h=\frac{1}{2}\big(1+\sqrt{1+m^2}\big).
\label{eq:gmy_49}
\end{equation}
Here, $h$ denotes the conformal weight of the scalar field. Since the one-loop partition function is defined as $Z_{\rm scalar}^{\rm 1\text{-}loop}=(\det\Delta)^{-1/2}$, its logarithm is given by
\begin{equation}
\log Z_{\rm scalar}^{\rm 1\text{-}loop}
=
\sum_{n=1}^\infty \frac{|q|^{2nh}}{n\,|1-q^n|^2}.
\label{eq:gmy_logZ}
\end{equation}
Exponentiating this sum yields the standard multi-particle product form \cite{GMY}
\begin{equation}
Z_{\rm scalar}^{\rm 1\text{-}loop}(\tau,\bar\tau)
=
\prod_{\ell,\bar\ell=0}^\infty \frac{1}{1-q^{\ell+h}\bar q^{\bar\ell+h}}.
\label{eq:gmy_410}
\end{equation}

The Wilson spool framework \cite{WilsonSpool,BourneFlissKnighton} reformulates these bulk functional determinants in terms of gauge-theory variables. In three dimensions, gravity can be formulated as a Chern-Simons gauge theory with  connections $A_L$ and $A_R$. For a smooth hyperbolic quotient $M\simeq \mathbb{H}^3/\Gamma$, the partition function is written as
\begin{equation}
\log Z_{\Delta,s}^{M}[g_{\mu\nu}]
=
W_\Gamma[A_L,A_R],
\label{eq:fliss_31}
\end{equation}
where the spool operator $W_\Gamma$ is a sum over the conjugacy classes $[\gamma]^+$ of the quotient group $\Gamma$ and over a specified set of lowest-weight representations $R$:
\begin{equation}
W_\Gamma
=
\sum_{[\gamma]^+}\;
\sum_{R\in R^{\rm LW}_{\Delta,s}}
\frac{1}{n_\gamma}
\Big[\Tr_{R_L}\,{\cal P}\exp\!\Big(\oint_\gamma A_L\Big)\Big]\,
\Big[\Tr_{R_R}\,{\cal P}\exp\!\Big(-\oint_\gamma A_R\Big)\Big].
\label{eq:fliss_32}
\end{equation}
Here, $n_\gamma$ counts the number of times a path winds around a primitive geodesic. In the thermal case where $\Gamma=\mathbb{Z}$, there is only one primitive element $\gamma_0$, corresponding to the thermal circle. The spool then reduces to a sum over integer windings $n\ge 1$, reproducing Eq.~\eqref{eq:gmy_logZ}. Each term in this sum arises from a pole contribution in the relevant integral, which emerges as an intermediate step in evaluating Eq.~\eqref{eq:fliss_32}. The terms in the sum~\eqref{eq:gmy_logZ} are thus interpreted as contributions from multi-winding loops, with $n$ ranging from $1$ to $\infty$. Our goal is to unpack this multi-winding sum as a sum over individual states of the scalar field in its Fock space at a fixed particle number $N$, and to reinterpret it as a sum over single-winding loops in higher representations, rather than as multi-winding loops in a fixed lowest-weight representation. Eventually we will argue that the states of matter fields in $AdS_3$ can be represented as single-loop Wilson lines in appropriate representations.

For a fixed $N$, all loops with winding numbers ranging from $1$ up to $n = N$ contribute to the fixed-$N$ sector. To recast these contributions as single-loop Wilson lines in appropriate representations, we must systematically unpack this mixing.

We now employ the dual boundary CFT language to rewrite Eq.~\eqref{eq:gmy_logZ} in terms of the global character of a CFT primary. Let ${\cal O}$ be the primary operator of weights $(h,\bar{h})=(h,h)$ dual to the bulk scalar field. The global $SL(2,\mathbb{R})_L\times SL(2,\mathbb{R})_R$ descendants generated by the action of the Virasoro modes $L_{-1}$ and $\bar L_{-1}$ span the single-particle Hilbert space
\begin{equation}
{\cal H}_1 \equiv {\cal V}_h\otimes \bar{\cal V}_h
=
\mathrm{span}\left\{\,L_{-1}^{\ell}\bar L_{-1}^{\bar\ell}\,|{\cal O}\rangle\ :\ \ell,\bar\ell\ge0\,\right\}.
\label{eq:H1_def}
\end{equation}
The global character of this module serves as our seed character:
\begin{equation}
\chi_1(q,\bar q)\equiv \Tr_{{\cal H}_1} q^{L_0}\bar q^{\bar L_0} = \frac{q^h\bar q^h}{(1-q)(1-\bar q)}.
\label{eq:seed_character}
\end{equation}
Using this seed character, the partition function $\log Z$ in Eq.~\eqref{eq:gmy_logZ} can be written compactly as
\begin{equation}
\log Z_{\rm scalar}^{\rm 1\text{-}loop}
=
\sum_{n=1}^\infty \frac{1}{n}\,\chi_1(q^n,\bar q^n).
\label{eq:logZ_as_PE}
\end{equation}
Exponentiating this relation defines the full partition function via the plethystic exponential ($\mathrm{PE}$) operator, which acts on a function of two variables as $\mathrm{PE}[f(q,\bar{q})] \equiv \exp\left(\sum_{n=1}^\infty \frac{1}{n} f(q^n, \bar{q}^n)\right)$. This gives:
\begin{equation}
Z_{\rm scalar}^{\rm 1\text{-}loop}
=
\exp\!\Big(\sum_{n=1}^\infty \frac{1}{n}\,\chi_1(q^n,\bar q^n)\Big) = \mathrm{PE}\big[\chi_1(q,\bar q)\big].
\label{eq:Z_as_PE}
\end{equation}
Physically, this expression represents the grand-canonical partition function of a bosonic Fock space built entirely from the single-particle module ${\cal H}_1$.

To keep track of the contribution of the fixed-$N$ sector to this partition function, we introduce a chemical potential, or fugacity, $t$. We define the refined partition function as
\begin{equation}
Z(t)\equiv \prod_{\ell,\bar\ell=0}^\infty \frac{1}{1-t\,q^{\ell+h}\bar q^{\bar\ell+h}}
=
\exp\!\Big(\sum_{n=1}^\infty \frac{t^n}{n}\,\chi_1(q^n,\bar q^n)\Big).
\label{eq:Zt_def}
\end{equation}
Expanding $Z(t)$ as a formal power series in $t$ allows us to isolate the fixed-$N$ multi-particle characters $\chi_N$:
\begin{equation}
Z(t)=\sum_{N\ge0} t^N\,\chi_N(q,\bar q),
\qquad \text{where} \qquad
\chi_N(q,\bar q)=\Tr_{\mathrm{Sym}^N({\cal H}_1)} q^{L_0}\bar q^{\bar L_0}.
\label{eq:chiN_def}
\end{equation}

To see how these multi-particle characters are constructed explicitly from the seed character, we expand the exponential form of $Z(t)$ :
\begin{align}
Z(t) &= \exp\left( t \chi_1(q,\bar{q}) + \frac{t^2}{2} \chi_1(q^2,\bar{q}^2) + \frac{t^3}{3} \chi_1(q^3,\bar{q}^3) + \mathcal{O}(t^4) \right) \nonumber \\
&= 1 + \left( t \chi_1 + \frac{t^2}{2} \chi_1(q^2) + \frac{t^3}{3} \chi_1(q^3) \right) + \frac{1}{2!}\left( t \chi_1 + \frac{t^2}{2} \chi_1(q^2) \right)^2 + \frac{1}{3!}\left( t \chi_1 \right)^3 + \mathcal{O}(t^4) \nonumber \\
&= 1 + t \, \chi_1(q,\bar{q}) + t^2 \left[ \frac{1}{2}\chi_1(q^2,\bar{q}^2) + \frac{1}{2}\left(\chi_1(q,\bar{q})\right)^2 \right] \nonumber \\
& \quad + t^3 \left[ \frac{1}{3}\chi_1(q^3,\bar{q}^3) + \frac{1}{2}\chi_1(q,\bar{q})\chi_1(q^2,\bar{q}^2) + \frac{1}{6}\left(\chi_1(q,\bar{q})\right)^3 \right] + \mathcal{O}(t^4).
\label{eq:Zt_expansion}
\end{align}
Matching the coefficients of $t^N$ between Eq.~\eqref{eq:Zt_expansion} and Eq.~\eqref{eq:chiN_def} yields explicit expressions for the multi-particle characters:
\begin{itemize}
    \item \textbf{For $N=1$:} The single-particle character matches the seed character itself,
    \begin{equation}
    \chi_1(q,\bar{q}) = \frac{q^h\bar{q}^h}{(1-q)(1-\bar{q})}.
    \label{eq:chi1_example}
    \end{equation}
    \item \textbf{For $N=2$:} The two-particle character corresponds to the symmetric tensor product $\mathrm{Sym}^2({\cal H}_1)$, given by
    \begin{equation}
    \chi_2(q,\bar{q}) = \frac{1}{2}\left( \left[\chi_1(q,\bar{q})\right]^2 + \chi_1(q^2,\bar{q}^2) \right).
    \label{eq:chi2_example}
    \end{equation}
    \item \textbf{For $N=3$:} The three-particle character corresponds to $\mathrm{Sym}^3({\cal H}_1)$ and takes the form
    \begin{equation}
    \chi_3(q,\bar{q}) = \frac{1}{6}\left( \left[\chi_1(q,\bar{q})\right]^3 + 3\chi_1(q,\bar{q})\chi_1(q^2,\bar{q}^2) + 2\chi_1(q^3,\bar{q}^3) \right).
    \label{eq:chi3_example}
    \end{equation}
\end{itemize}

As mentioned earlier, for a fixed $N$-sector there is a mixing of multi-winding contributions, and our goal is to systematically unpack these $N$-particle characters. We will also write down the general relationship between the $N$-sector character $\chi_N$ and the multi-winding contributions using Schur--Weyl/Frobenius technology, see Appendix \ref{sec:schur_weyl_review}.

\subsection{From $\chi_N$ to Multi-Trace Primaries}
\label{subsec:fixedN_to_primaries}

A Wilson loop in an appropriate representation encodes the global character of the corresponding primary. To express $\chi_N$ as a sum over single-winding loops, we must identify the primary states in the Fock space at each level $(h_p, \bar{h}_p)$: summing the global characters of these primaries then reproduces the fixed-$N$ contribution $\chi_N$. The characters $\chi_N$ count all states in the bosonic $N$-particle sector, including global descendants generated by $L_{-1}$ and $\bar{L}_{-1}$.

In this section, we extract the levels and multiplicities of these primaries from $\chi_N$. To do so, we strip off the global descendants, which is straightforward since the descendant structure is universal. What remains is the purely primary contribution. The $\mathrm{AdS}_3/\mathrm{CFT}_2$ correspondence further tells us that, in the $c \to \infty$ limit, the multi-particle states of a scalar field are dual to multi-trace primary operators in the boundary CFT. The spectrum of these multi-trace primaries --- their levels and multiplicities --- can therefore be read off explicitly.

A global primary of weights $(h_P,\bar h_P)$ contributes the universal global module character
\begin{equation}
\chi_{h_P,\bar h_P}(q,\bar q) = \frac{q^{h_P}\bar q^{\bar h_P}}{(1-q)(1-\bar q)}.
\label{eq:global_character}
\end{equation}
Therefore, the generating function that counts primaries (with multiplicity) is obtained by removing the universal $(L_{-1},\bar L_{-1})$ towers:
\begin{equation}
P_N(q,\bar q)\defi (1-q)(1-\bar q)\,\chi_N(q,\bar q) = \sum_{\text{primaries }P\ \text{in }N\text{-sector}} \mathrm{mult}(P)\,q^{h_P}\bar q^{\bar h_P}.
\label{eq:PN_def}
\end{equation}
By construction, $\chi_N$ is the character of the $N$-particle operator space ${\cal A}_N\defi \mathrm{Sym}^N({\cal H}_1)$, while $P_N$ is the character of the primary subspace after quotienting by total derivatives. Equivalently, in terms of operators, ${\cal A}_N$ is spanned by normal-ordered products of $N$ copies of ${\cal O}$ and its derivatives, $:\!\prod_{i=1}^N \de^{m_i}\bar\de^{\bar m_i}{\cal O}\!:$, symmetrized under permutations of the identical factors. The coefficients in $P_N$ are exactly the number of linearly independent multi-trace primaries at each pair of weights.

We now work out $P_2$ and $P_3$ explicitly.

\subsubsection{Multi-Trace Primary Counting at $N=2$}
\label{subsec:N2_counting}

Starting from Eq.~\eqref{eq:chi2_example}, one finds
\begin{equation}
P_2(q,\bar q) = (1-q)(1-\bar q)\chi_2(q,\bar q) = (q\bar q)^{2h}\,\frac{1}{2}\left[ \frac{1}{(1-q)(1-\bar q)}+\frac{1}{(1+q)(1+\bar q)} \right].
\label{eq:P2_closed}
\end{equation}
Expanding $1/(1-q)=\sum_{m\ge0}q^m$ and $1/(1+q)=\sum_{m\ge0}(-1)^m q^m$ (and similarly for $\bar q$), we immediately read off
\begin{equation}
P_2(q,\bar q) = (q\bar q)^{2h}\!\!\sum_{m,\bar m\ge0} \frac{1+(-1)^{m+\bar m}}{2}\;q^m\bar q^{\bar m}.
\label{eq:P2_series}
\end{equation}
Thus the double-trace primary multiplicity is
\begin{equation}
d^{(2)}_{m,\bar m}=\frac{1+(-1)^{m+\bar m}}{2} =
\begin{cases}
1, & m\equiv \bar m\ (\mathrm{mod}\ 2),\\[2pt]
0, & m\not\equiv \bar m\ (\mathrm{mod}\ 2).
\end{cases}
\label{eq:d2_rule}
\end{equation}
Now, using Eq.~\eqref{eq:global_character}, Eq.~\eqref{eq:P2_series} is equivalent to
\begin{equation}
\chi_2(q,\bar q) = \sum_{\substack{m,\bar m\ge0\\ m\equiv \bar m\ (\mathrm{mod}\ 2)}} \chi_{2h+m,\;2h+\bar m}(q,\bar q) = \sum_{\substack{m,\bar m\ge0\\ m\equiv \bar m\ (\mathrm{mod}\ 2)}} \frac{q^{2h+m}\bar q^{2h+\bar m}}{(1-q)(1-\bar q)}.
\label{eq:N2_primary_decomp}
\end{equation}
The individual terms will be interpreted as characters of multi-trace primary Wilson loops.

The explicit count of these primaries in the CFT is as follows. A general double trace primary at the level $(m,\bar{m})$ can be written as $:\partial_1^{p}\partial_2^{q}\bar{\partial}_1^{r}\bar{\partial}_2^{s}\mathcal{O}(z_1,\bar{z}_1)\mathcal{O}(z_2,\bar{z}_2):$ with the understanding that $p+q=m$ and $r+s=\bar{m}$, together with the constraint that we should not act with the total derivative ($\partial_{\mathrm{tot}} = \partial_1 +\partial_2$, $\bar{\partial}_{\mathrm{tot}} = \bar{\partial}_1 +\bar{\partial}_2$) anywhere as it will generate global descendants. 

So for $(m,\bar{m}) = (0,0)$, the only possible values are $p=0, q=0, r=0, s=0$. The only primary is the bare product $:\mathcal{O}(z_1,\bar{z}_1)\mathcal{O}(z_2,\bar{z}_2):$ evaluated at a coincident point, which is trivially symmetric under particle exchange. 

For $(m,\bar{m}) = (1,0)$, the holomorphic derivatives must satisfy $p+q=1$, giving the choices $(p,q) = (1,0)$ or $(0,1)$, while the anti-holomorphic sector is restricted to $r=0, s=0$. This yields two raw states, $\partial_1 :\mathcal{O}\mathcal{O}:$ and $\partial_2 :\mathcal{O}\mathcal{O}:$. Removing the total derivative descendant $(\partial_1 + \partial_2):\mathcal{O}\mathcal{O}:$ leaves a single unique primary combination: $(\partial_1 - \partial_2):\mathcal{O}\mathcal{O}:$. However, this state is not bosonic because swapping the two identical particles ($1 \leftrightarrow 2$) interchanges the derivative powers ($p \leftrightarrow q$), mapping $\partial_1 - \partial_2 \rightarrow \partial_2 - \partial_1 = -(\partial_1 - \partial_2)$. Because it is anti-symmetric, it is projected out of the bosonic Hilbert space. A completely analogous argument holds for $(m,\bar{m}) = (0,1)$ where $(p,q)=(0,0)$ and $(r,s)=(1,0)$ or $(0,1)$, yielding the anti-symmetric primary $(\bar{\partial}_1 - \bar{\partial}_2):\mathcal{O}\mathcal{O}:$ that again gets projected out.

For $(m, \bar{m} ) =(1,1)$, the constraints allow four combinations of derivative powers since $(p,q)$ can be $(1,0)$ or $(0,1)$ and $(r,s)$ can be $(1,0)$ or $(0,1)$. This yields the four configurations for $(p,q,r,s)$: $(1,0,1,0)$, $(1,0,0,1)$, $(0,1,1,0)$, and $(0,1,0,1)$. Eliminating the states generated by total derivative actions ($\partial_{\mathrm{tot}}$ and $\bar{\partial}_{\mathrm{tot}}$) isolates the unique primary $(\partial_1 - \partial_2)(\bar{\partial}_1 - \bar{\partial}_2):\mathcal{O}\mathcal{O}:$. This state is bosonic because swapping the two slots exchanges both sets of indices simultaneously ($p \leftrightarrow q$ and $r \leftrightarrow s$). This transforms the operator into $(\partial_2 - \partial_1)(\bar{\partial}_2 - \bar{\partial}_1)$, and the two resulting sign flips cancel out ($(-1) \times (-1) = +1$) to leave the primary completely symmetric. 

This method of removing the descendant direction gives the correct count of primaries at any $N$. But the explicit operators it produces are a priori only {\em representatives} of the bosonic invariant space and are not necessarily annihilated by the global raising operators $L_1^{\rm tot},\bar L_1^{\rm tot}$, as a true global primary must be. However, for the $N=2$ case constructed above, it so happens that the operators written above are indeed true global primaries. We return to this point in Appendix~\ref{sec:honest_primaries}, where the primary condition is imposed explicitly for $N=2$ and $N=3$.

A general strategy for the construction of the (symmetric space of) bosonic multi-trace operators for a general $N$-particle sector, together with some explicit examples for $N=2,3$, is presented in Appendix~\ref{sec:Multi_trace_primary_construction} and Appendix~\ref{sec:honest_primaries}.

\subsubsection{Multi-Trace Primary Counting at $N=3$}
\label{subsec:N3_counting}

Starting from Eq.~\eqref{eq:chi3_example}, one finds the compact expression
\begin{equation}
P_3(q,\bar q) = (q\bar q)^{3h}\left[ \frac{1}{6}\frac{1}{(1-q)^2(1-\bar q)^2} +\frac{1}{2}\frac{1}{(1-q^2)(1-\bar q^2)} +\frac{1}{3}\frac{(1-q)(1-\bar q)}{(1-q^3)(1-\bar q^3)} \right].
\label{eq:P3_closed}
\end{equation}
Writing
\begin{equation}
P_3(q,\bar q)=(q\bar q)^{3h}\sum_{m,\bar m\ge0} d^{(3)}_{m,\bar m}\,q^m\bar q^{\bar m},
\label{eq:P3_series_def}
\end{equation}
The coefficients $d^{(3)}_{m,\bar m} \in \mathbb{Z}_{\ge 0}$ represent the multiplicities of triple-trace primaries at weights $(3h+m, 3h+\bar m)$. While $d^{(3)}_{m, \bar{m}}$ could be derived by expanding the right-hand side of Eq.~\ref{eq:P3_closed} in powers of $q^m \bar{q}^{\bar{m}}$-analogous to the calculation for $N=2$-we instead use group theory to derive a general expression for arbitrary $N$ in Appendix~\ref{sec:Multi_trace_primary_construction}. For $N=3$, the multiplicity is
\begin{equation}
d^{(3)}_{m,\bar m} = \frac{(m+1)(\bar m+1)}{6} + \frac{1}{2} \mathbf{1}_{m \text{ even}} \mathbf{1}_{\bar m \text{ even}} + \frac{1}{3} c_m c_{\bar m},
\label{eq:d3_formula}
\end{equation}

where
\begin{equation}
c_m=
\begin{cases}
1, & m\equiv 0\ (\mathrm{mod}\ 3),\\
-1, & m\equiv 1\ (\mathrm{mod}\ 3),\\
0, & m\equiv 2\ (\mathrm{mod}\ 3).
\end{cases}
\label{eq:cm_def}
\end{equation}

Combining \eqref{eq:P3_series_def} with Eq.~\eqref{eq:global_character} yields
\begin{equation}
\chi_3(q,\bar q) = \sum_{m,\bar m\ge0} d^{(3)}_{m,\bar m}\,\chi_{3h+m,\;3h+\bar m}(q,\bar q) = \sum_{m,\bar m\ge0} d^{(3)}_{m,\bar m}\,\frac{q^{3h+m}\bar q^{3h+\bar m}}{(1-q)(1-\bar q)}.
\label{eq:N3_primary_decomp}
\end{equation}
The expression for the multiplicity at general $N$ is presented in \eqref{count}.

\subsection{Multi-trace Wilson loops: ``One Wilson Loop per Primary''}
\label{subsec:multitrace_narrative}

In the Chern--Simons/holonomy description underlying the Wilson-spool 
construction, the thermal $\mathrm{AdS}_3$ background fixes the holonomies 
of the gauge connections $A_L$ and $A_R$ around the thermal cycle. 
Specifically, in the lowest-weight (discrete-series) representation 
$\mathcal{D}^+_{h_P}$ of $SL(2,\mathbb{R})_L$, the eigenvalues of 
$\mathcal{P}\exp\!\big(\oint_{\gamma_0} A_L\big)$ are $\{q^{h_P+\ell}\}_{\ell\ge0}$, 
so the trace evaluates to $q^{h_P}/(1-q)$. An analogous computation for 
$\mathcal{P}\exp\!\big(-\oint_{\gamma_0} A_R\big)$ in the representation 
$\mathcal{D}^+_{\bar h_P}$ of $SL(2,\mathbb{R})_R$ gives 
$\bar q^{\bar h_P}/(1-\bar q)$. A Wilson loop wrapping the thermal cycle 
once in the representation $(h_P,\bar h_P)$ therefore evaluates to the 
corresponding global character. Denoting this single-winding loop by 
$W_{(h_P,\bar h_P)}$, we have
\begin{equation}
W_{(h_P,\bar h_P)}(q,\bar q) = \chi_{h_P,\bar h_P}(q,\bar q) = \frac{q^{h_P}\bar q^{\bar h_P}}{(1-q)(1-\bar q)}.
\label{eq:wilsonloop_equals_character}
\end{equation}
This identification is the key bridge between the spool framework and 
the primary decomposition derived above. The fixed-$N$ decompositions 
\eqref{eq:N2_primary_decomp} and \eqref{eq:N3_primary_decomp} can then 
be read as a rewriting of $\chi_N$ as a sum of single-winding Wilson 
loops, one per independent multi-trace primary counted with multiplicity:
\begin{equation}
\chi_2(q,\bar q) = \sum_{\substack{m,\bar m\ge0\\ m\equiv \bar m\ (\mathrm{mod}\ 2)}} W_{(2h+m,\;2h+\bar m)}(q,\bar q),
\label{eq:N2_wilson_sum}
\end{equation}
\begin{equation}
\chi_3(q,\bar q) = \sum_{m,\bar m\ge0} d^{(3)}_{m,\bar m}\,W_{(3h+m,\;3h+\bar m)}(q,\bar q),
\label{eq:N3_wilson_sum}
\end{equation}
For a general $N$-particle sector:

\begin{equation}
\chi_N(q,\bar q) = \sum_{m,\bar m\ge0} d^{(N)}_{m,\bar m}\,W_{(Nh+m,\;Nh+\bar m)}(q,\bar q),
\label{eq:NN_wilson_sum}
\end{equation}
with $d^{(N)}_{m,\bar m}$ provided in \eqref{count}.

This result realizes the picture advocated 
in Section~\ref{subsec:gmy_fliss_review}: the states of the $N$-particle 
Fock space are precisely accounted for by single-winding Wilson lines in appropriate 
higher representations, built from the seed primary  of weight $h$. The lesson for us, is broader: we will view this as a hint that in the TQFT/Wilson-network language, weakly coupled local fields can be understood via the collective physics of such towers of Wilson line primaries.

\section{Boundary Graviton Dressing: Global to Virasoro Characters}
\label{subsec:virasoro_dressing}

Up to this point we have organized the scalar one-loop determinant into a sum over \emph{global}
$SL(2)_L\times SL(2)_R$ modules.  Concretely, the one-particle space ${\cal H}_1$ in
Eq.~\eqref{eq:H1_def} was built only from the global descendants generated by $L_{-1}$ and
$\bar L_{-1}$, and the basic building block that appears throughout is the global primary character
$\chi_{h_P,\bar h_P}(q,\bar q)$ defined in Eq.~\eqref{eq:global_character}.  This is exactly what one
expects from a matter determinant computed on a \emph{fixed} thermal $\mathrm{AdS}_3$ background: it
counts the bulk scalar excitations and their $SL(2)$ descendants, but it does not by itself account
for the additional Virasoro descendant structure associated with boundary gravitons. In this section, we will note that the boundary gravitons have precisely the structure needed to turn the global characters into a Virasoro character -- this is a necessary ingredient in interpreting these results as pieces of a CFT partition function.

In $\mathrm{AdS}_3/\mathrm{CFT}_2$, the asymptotic symmetry algebra is Virasoro, and the universal
``boundary graviton'' excitations are generated by the modes $L_{-n}$ with $n\ge2$ (and similarly in
the right-moving sector).  The corresponding chiral vacuum-module character is therefore
\begin{equation}
\chi^{\rm Vir}_{\rm vac}(q)\defi \prod_{n=2}^{\infty}\frac{1}{1-q^n},
\qquad
\bar\chi^{\rm Vir}_{\rm vac}(\bar q)\defi \prod_{n=2}^{\infty}\frac{1}{1-\bar q^{\,n}}.
\label{eq:vir_vac_char}
\end{equation}
The product starts at $n=2$ because $L_{-1}|0\rangle=0$ in the vacuum module. We suppress the
universal Casimir-energy prefactors, since our focus here is on the descendant counting and the
organization into modules.

For a non-vacuum primary state $|P\rangle$ of left weight $h_P$, the full Virasoro Verma module
contains descendants created by all $L_{-n}$ with $n\ge1$.  Its chiral character may be written as
\begin{equation}
\chi^{\rm Vir}_{h_P}(q)
\defi
q^{h_P}\prod_{n=1}^{\infty}\frac{1}{1-q^n}
=
\frac{q^{h_P}}{1-q}\,\chi^{\rm Vir}_{\rm vac}(q),
\label{eq:vir_primary_char}
\end{equation}
and similarly $\bar\chi^{\rm Vir}_{\bar h_P}(\bar q)=\bar q^{\,\bar h_P}\prod_{n\ge1}(1-\bar q^{\,n})^{-1}$.
The factor $(1-q)^{-1}$ is precisely what is \emph{absent} from the vacuum character
\eqref{eq:vir_vac_char}, and it is supplied by the level-one $SL(2)$ descendant tower already present
in the global module.

Combining the left- and right-moving sectors, the full non-chiral Virasoro character of a primary
with weights $(h_P,\bar h_P)$ is
\begin{equation}
\chi^{\rm Vir}_{h_P,\bar h_P}(q,\bar q)
\defi
\chi^{\rm Vir}_{h_P}(q)\,\bar\chi^{\rm Vir}_{\bar h_P}(\bar q)
=
\chi^{\rm Vir}_{\rm vac}(q)\,\bar\chi^{\rm Vir}_{\rm vac}(\bar q)\;
\frac{q^{h_P}\bar q^{\bar h_P}}{(1-q)(1-\bar q)}.
\label{eq:vir_dressing_raw}
\end{equation}
Using Eq.~\eqref{eq:global_character}, we can package this as the simple ``dressing'' relation
\begin{equation}
\chi^{\rm Vir}_{h_P,\bar h_P}(q,\bar q)
=
\chi^{\rm Vir}_{\rm vac}(q)\,\bar\chi^{\rm Vir}_{\rm vac}(\bar q)\;
\chi_{h_P,\bar h_P}(q,\bar q),
\label{eq:vir_dressing}
\end{equation}
where $\chi_{h_P,\bar h_P}(q,\bar q)$ is the \emph{global} module character used throughout
Secs.~\ref{subsec:multitrace_narrative}.  In words:
our $SL(2)_L\times SL(2)_R$ primary bookkeeping (e.g.\ the fixed-$N$ decomposition into primaries and
multiplicities) together with the boundary
gravitons provide a universal multiplicative dressing of each module.

In fact, since the dressing factor $\chi^{\rm Vir}_{\rm vac}(q)\bar\chi^{\rm Vir}_{\rm vac}(\bar q)$ is
independent of the scalar multi-particle sector, it multiplies the entire refined grand canonical
partition function \eqref{eq:Zt_def}:
\begin{equation}
Z^{\rm dressed}(t)
\defi
\chi^{\rm Vir}_{\rm vac}(q)\,\bar\chi^{\rm Vir}_{\rm vac}(\bar q)\;Z(t)
=
\sum_{N\ge0} t^N\,\chi_N^{\rm dressed}(q,\bar q),
\qquad
\chi_N^{\rm dressed}
=
\chi^{\rm Vir}_{\rm vac}(q)\,\bar\chi^{\rm Vir}_{\rm vac}(\bar q)\;\chi_N.
\label{eq:dressed_Zt}
\end{equation}
Equivalently, in the ``one module per primary'' organization of Sec.~\ref{subsec:multitrace_narrative},
each global character appearing in the fixed-$N$ decompositions is promoted to the corresponding
Virasoro character:
\begin{equation}
\chi_N^{\rm dressed}(q,\bar q)
=
\sum_{\text{primaries }P\ \text{in }N\text{-sector}}
\mathrm{mult}(P)\,\chi^{\rm Vir}_{h_P,\bar h_P}(q,\bar q),
\qquad
\chi^{\rm Vir}_{h_P,\bar h_P}\ \text{as in Eq.~\eqref{eq:vir_dressing_raw}}.
\label{eq:dressed_fixedN_decomp}
\end{equation}
This makes explicit how the $SL(2)$ primary structure extracted in Eqs.~\eqref{eq:PN_def},
\eqref{eq:N2_primary_decomp} and \eqref{eq:N3_primary_decomp} fits together with boundary-graviton
dressing to yield full Virasoro modules: $L_{-1}$ (and $\bar L_{-1}$) are already accounted for in the
global characters, while the genuinely new Virasoro descendants $L_{-n\ge2}$ are universal and
captured by the vacuum factor \eqref{eq:vir_vac_char}.

\subsection{Alternative Book-keeping: Virasoro Characters from Orbits}

In the holonomy picture where a single-winding Wilson loop evaluates to the global character
\eqref{eq:wilsonloop_equals_character}, the dressed module \eqref{eq:vir_dressing} may be viewed as a
``Virasoro-dressed'' loop,
\begin{equation}
\widetilde W_{(h_P,\bar h_P)}(\tau,\bar\tau)
\defi
\chi^{\rm Vir}_{\rm vac}(q)\,\bar\chi^{\rm Vir}_{\rm vac}(\bar q)\;
W_{(h_P,\bar h_P)}(\tau,\bar\tau),
\label{eq:dressed_wilson}
\end{equation}
so that each term in the fixed-$N$ Wilson-loop sums (e.g.\ \eqref{eq:N2_wilson_sum}--\eqref{eq:N3_wilson_sum})
is promoted from a global module contribution to a full Virasoro-module contribution. 

In fact there is alternative, more direct, way to compute these Virasoro-dressed loops/characters from the bulk. The idea here is to reduce the bulk Chern-Simons action in the presence of a Wilson line source to a boundary chiral WZW model and then to an Alekseev-Shatashvili orbit action \cite{AS}. This allows an exact path integral evaluation on the torus \cite{Pradipta,CotlerJensen, Chua-Jiang} and the result is directly the character corresponding to the primary label of the Wilson loop.

The lesson from either of these organizations is that any time a multi-trace tower of primaries is part of the spectrum of a CFT, it leads to a tower of Wilson line labels in the bulk whose contribution to the CFT partition function is precisely (the sum of) their chiral-anti-chiral paired Virasoro characters. This ties in naturally with the Wilson network definition of the holographic CFT via its Teichm\"uller/Virasoro TQFT skeleton \cite{KashaevQT, PonsotTeschner1,PonsotTeschner2, CEZ1, Vishal}. 

\section{BTZ Handlebody: The Defect-Twined Character}
\label{sec:heavy_light_cycle_exchange}

We now turn to the interpretation of the Wilson spool in the BTZ handlebody.
We will first isolate a simpler and more universal object: the
defect-twined character obtained by placing a closed light Wilson line around
a heavy Wilson-line source.  This object is the natural monodromy observable of
a light primary in the background of a fixed heavy primary. 

This a natural object from the TQFT skeleton perspective of \cite{Vishal}. In the Teichm\"uller (Virasoro) TQFT description, a heavy
primary is represented by a Wilson line along the core of the thermal solid
torus, and the universal gauge-invariant way to probe such a source is to
measure the holonomy around it.  This is done by a second Wilson line on a
meridian linking the heavy core.

It is noteworthy that this basic ingredient already carries a hint of the cycle-exchange present in the BTZ handlebody: the
linking loop that detects the monodromy of the heavy Wilson-line source is necessarily spatial. We will see that the
cycle-exchanged BTZ interpretation emerges, once this fixed-heavy-sector
observable is summed over the heavy spectrum with the Virasoro Plancherel (essentially Cardy) density.

\subsection{The Modular Kernel and BTZ}
\label{subsec:btz_no_probe_kernel}

We first recall the no-probe statement.  For $c>1$ write
\[
c=1+6Q^2
\]
and label a non-degenerate Virasoro representation by a real momentum
$P\geq 0$, with
\[
h(P)=\frac{Q^2}{4}+P^2 .
\]
The corresponding chiral character is
\begin{equation}
\widehat\chi_P(\tau)
=
\Tr_{{\cal V}_P}
\left(q^{L_0-\frac{c}{24}}\right)
=
\frac{q^{P^2}}{\eta(\tau)}
\qquad
q=e^{2\pi i\tau}.
\label{eq:btz_cont_char_def}
\end{equation}
The modular transform of a character is written\footnote{The explicit forms of the S-kernels are well-known, but we will not write them because we will not need them.} as
\begin{equation}
\widehat\chi_A\!\left(-\frac{1}{\tau}\right)
=
\int_0^\infty dP\,
S_{AP}(c)\,
\widehat\chi_P(\tau).
\label{eq:btz_S_kernel_general}
\end{equation}
For a non-degenerate label $A$, the kernel is the usual Virasoro continuum
kernel.  The vacuum row is special and gives the Cardy/Plancherel density,
\begin{equation}
\widehat\chi_{\rm vac}\!\left(-\frac{1}{\tau}\right)
=
\int_0^\infty dP\,
S_{0P}(c)\,
\widehat\chi_P(\tau).
\label{eq:BTZ_vac_row}
\end{equation}
This is the chiral version of the statement that the smooth BTZ handlebody is
represented, in the original thermal-AdS channel, by a universal heavy-primary
density.  The full non-chiral answer contains the corresponding
antiholomorphic factor. In the bulk language, this translates to the statement that the BTZ partition function is a stand-in for the contributions from an ensemble of heavy Wilson line primaries (and a subleading contribution from the boundary graviton descendants) \cite{Pradipta}. 

\subsection{The Defect-Twined Character of a Heavy Primary}
\label{subsec:defect_twined_character}

Now fix a heavy primary, with Virasoro momentum $P$.  In the TQFT
description this state is represented by a Wilson line labelled by $P$ along
the core of the solid torus.  A light primary label $a$ defines a topological
defect or Verlinde line $\mathsf D_a$.  The natural closed-worldline probe of
the heavy source is the line $\mathsf D_a$ wrapped on a meridian linking the
core Wilson line.

The corresponding CFT object is not an ordinary thermal trace of local light
excitations.  It is the defect-twined character of the heavy module:
\begin{equation}
{\cal Z}^{\rm tw}_{a|P}(\tau)
\defi
\Tr_{{\cal V}_{P}}
\left(
\mathsf D_a\,
q^{L_0-\frac{c}{24}}
\right).
\label{eq:defect_twined_character_def}
\end{equation}
Because $\mathsf D_a$ is topological, it commutes with the Virasoro algebra and
therefore acts by a scalar on an irreducible Virasoro family (via Schur's lemma).  In the standard
Verlinde normalization,
\begin{equation}
\mathsf D_a\big|_{{\cal V}_P}
=
\frac{S_{aP}}{S_{0P}}\,
\mathbf 1_{{\cal V}_P}.
\label{eq:verlinde_line_eigenvalue}
\end{equation}
Hence
\begin{equation}
{\cal Z}^{\rm tw}_{a|P}(\tau)
=
\frac{S_{aP}}{S_{0P}}\,
\widehat\chi_{P}(\tau).
\label{eq:defect_twined_character_result}
\end{equation}

This is the precise sense in which we attach a light primary to a fixed heavy
primary in this section. We are computing the monodromy
observable of the light representation around the heavy Wilson-line source. In a TQFT, this a natural partition function one can associate to the intuition ``light primary on a heavy primary". 

The same formula can be derived or motivated from a standard surgery interpretation.  A solid torus with
a core line labeled by $P$ prepares the state $|P\rangle$ in the torus Hilbert
space.  Gluing by the modular $S$ transformation turns two core lines into a
Hopf link, so
\[
Z\!\left(S^3,\mathrm{Hopf}(a,P)\right)=S_{aP}
\qquad
Z\!\left(S^3,\text{unknot }P\right)=S_{0P}.
\]
The normalized linking amplitude is therefore $S_{aP}/S_{0P}$.  In an
abelian Chern--Simons theory this ratio is just the Aharonov--Bohm phase of
one charge encircling another.  The Virasoro/Teichm\"uller formula is the
corresponding continuum, non-abelian monodromy eigenvalue
\cite{CEZ1,CEZ2,Verlinde,PetkovaZuber,BuicanGromovDefects,WittenJones}.

\subsection{Summing the Heavy Sectors}
\label{subsec:heavy_sum_defect_twined}

We now sum the fixed-heavy-sector observable over the universal heavy density.
Using the vacuum-row kernel \eqref{eq:BTZ_vac_row}, the defect-twined heavy
sum is
\begin{equation}
{\cal Z}^{\rm tw}_{a}(\tau)
=
\int_0^\infty dP\,
S_{0P}(c)\,
\frac{S_{aP}(c)}{S_{0P}(c)}\,
\widehat\chi_P(\tau).
\label{eq:defect_twined_heavy_sum}
\end{equation}
The factor $S_{0P}$ is the ordinary heavy-primary density supplied by the
BTZ/Cardy saddle.  The ratio $S_{aP}/S_{0P}$ is the monodromy eigenvalue of
the light defect around a heavy primary with momentum $P$.  Their product is
therefore the unnormalized defect-weighted heavy density:
\[
S_{0P}\,
\frac{S_{aP}}{S_{0P}}
=
S_{aP}.
\]
Consequently
\begin{equation}
{\cal Z}^{\rm tw}_{a}(\tau)
=
\int_0^\infty dP\,
S_{aP}(c)\,
\widehat\chi_P(\tau)
=
\widehat\chi_a\!\left(-\frac{1}{\tau}\right).
\label{eq:defect_twined_cycle_exchange}
\end{equation}
This is the cycle-exchange statement relevant for the spool.  In the original
thermal-AdS channel, the light line is a meridian linking the heavy Wilson-line
source.  After summing over heavy labels with the modular density, the same
object is reorganized as the dual-channel character of the light primary.
This is the sense in which the light loop becomes the spatial loop around the
non-contractible cycle of the BTZ handlebody. A sum over a tower of such primaries (via say, plethystic exponentiation) is the one-loop determinant aka Wilson spool \cite{WilsonSpool,BourneFlissKnighton} -- the structure of that step is the same as in thermal AdS$_3$, so we will not repeat it.  We also note that the full answer involves both chiral halves, as it did in AdS$_3$.

Note that {\em if} one grants that the underlying UV-complete bulk description of AdS$_3$/CFT$_2$ is a TQFT with Wilson lines, the result (and the calculation from the twined character) is natural: this was our motivation is using the TQFT language to derive the eigenvalue of the Verlinde defect. But from a bulk EFT/gravity point of view, it is completely mysterious. This calculation can therefore be viewed as giving evidence that the TQFT skeleton \cite{Vishal} is the fundamental approach to AdS$_3$/CFT$_2$.

\subsection{A Modular Bootstrap for Twined Partition Functions?}
\label{subsec:scope_defect_twined}

In this section, our goal was to give a natural understanding of the origin of the one loop determinant, on the BTZ saddle. We did that in terms of (a plethystic tower of) light Wilson line primaries on a dense family of heavy Wilson line primaries, in a TQFT. The twined character of the light primary on the heavy primary, played a natural role.

There is nevertheless a stronger and more microscopic statement one might try to derive from local CFT data. One might try to justify (the usually {\em assumed}) statement that a light probe on the BTZ handlebody, is a {\em meaningful} CFT construct. This would justify the assumption that probes see a smooth horizon. Depending on the resolution of the question, this will require ideas like ETH \cite{Deutsch, Srednicki} and beyond. It will be interesting to formulate these from the TQFT perspective.

A less ambitious possibility is to study the modular covariance properties of the twined partition function. In holographic CFTs where light probes on smooth black holes are expected to make sense, a natural guess is that the analogue (of vacuum dominance in the dual channel for the partition function) would be dominance of the light primary character in the dual channel for the twined partition function. 

We will not develop these ideas in this paper, but there is an interesting hint regarding this in the functional form of $S_{aP}$. From its explicit form as the cosine kernel, it is clear that above the BTZ threshold it is an oscillatory function of either sign and clearly cannot have an interpretation as a meaningful density of states. But below the threshold, which is where the primaries are light and {\em have} a probe interpretation, the kernel is defined by analytic continuation and is positive definite.

\section*{Acknowledgments}

We thank Vaibhav Burman and Pradipta Pathak for discussions and Darsh Bhatt for collaborations at the early stages of this project. VG is supported by the Council of Scientific \& Industrial Research (CSIR) Fellowship No. 09/0079(22163)/2025-EMR-I. We thank ChatGPT/Claude for discussions and for pointers to directly useful references.

\appendix

\section{Multi-Trace Primaries and their Multiplicity}
\label{sec:Multi_trace_primary_construction}
To systematically construct the multi-trace primary operators at a general level $(m,\bar{m})$ above the base weight $(Nh,N\bar{h})$ in the $N$-particle sector, we analyze the structure of identical field products. In the context of the AdS/CFT, multi-particle states in the bulk are dual to local multi-trace operators in the CFT. We consider a configuration consisting of $N$ copies of a single-particle primary operator $\mathcal{O}(z,\bar{z})$ of conformal weight $(h,\bar h)=(h,h)$, since it is dual to a scalar field in the bulk.

A single copy of $\mathcal{O}$ generates its global descendants by acting with $L_{-1},\bar L_{-1}$. Correspondingly, a product of $N$ copies of $\mathcal O$ generates the full space of states at level $(m,\bar m)$ by acting with $L_{-1}^{(i)},\bar L_{-1}^{(i)}$ independently on each of the $N$ slots. In position space this operation is realized as independent holomorphic and anti-holomorphic derivatives acting on each field slot -- this is how the $N$-particle sector of the bulk Fock space is reproduced on the CFT side.

This allows us to write down a monomial configurations as
\begin{equation}
:\partial_1^{p_1}\dots\partial_N^{p_N}\bar{\partial}_1^{r_1}\dots\bar{\partial}_N^{r_N}\mathcal{O}(z_1,\bar{z}_1)\dots\mathcal{O}(z_N,\bar{z}_N):
\label{eq:raw-monomial-configuration}
\end{equation}
where the indices $i \in \{1, \dots, N\}$ label the specific operator slot that the derivative acts upon. The integers $p_i \ge 0$ and $r_i \ge 0$ denote the respective derivative powers, and the total conformal levels are restricted by $\sum_{i=1}^N p_i = m$ and $\sum_{i=1}^N r_i = \bar{m}$.

A genuine local operator insertion, dual to a definite multi-particle bulk state, must itself sit at a single point. We therefore start with the $N$ fields at separate points $z_1,\ldots,z_N$ and bring them together at the end of the construction. Since the operator product expansion makes the product singular as any two points collide, we normal order the product to subtract these short-distance singularities and leave a finite composite operator, as denoted by the colons in Eq.~\eqref{eq:raw-monomial-configuration}. The individual points $z_i$ are used only to build finite, well-defined derivative combinations before the coincident-point limit $z_i\to z$ is taken: it is this limit that turns the $N$-fold product into a single local multi-trace operator $[\mathcal O^N]_{m,\bar m}(z,\bar z)$ with a definite conformal weight, dual to a genuine $N$-particle state of the bulk field evaluated at a point. Although in what follows we will often write the $N$ copies of $\mathcal O$ as sitting at separate points $z_1,\ldots,z_N$ for convenience, it should be understood throughout that this coincident-point limit is taken at the end of the construction.

To frame this construction algebraically, let $W_N = \mathrm{span}\{\partial_1, \dots, \partial_N\}$ be the $N$-dimensional vector space of single-particle holomorphic derivatives, and similarly $\bar{W}_N = \mathrm{span}\{\bar{\partial}_1, \dots, \bar{\partial}_N\}$ for the anti-holomorphic derivatives. At fixed level $(m,\bar m)$, the full space of derivative monomials is the tensor product of symmetric powers
\begin{equation}
\mathrm{Sym}^m(W_N) \otimes \mathrm{Sym}^{\bar{m}}(\bar{W}_N) = \mathrm{span} \left\{ \prod_{i=1}^N \partial_i^{p_i} \bar{\partial}_i^{r_i} \ \Bigg| \sum_{i=1}^N p_i = m, \sum_{i=1}^N r_i = \bar{m} \right\}~.
\end{equation}
These raw monomials, obtained by simply distributing derivatives among the $N$ copies of $\mathcal{O}$, as in Eq.~\eqref{eq:raw-monomial-configuration}, do not directly correspond to physical primary operators due to the following two conditions that must be imposed.

The first condition is bosonic symmetry. The $N$ copies of $\mathcal{O}$ are identical, and since we are dealing with a scalar field in the bulk, physical states must be invariant under the symmetric group $S_N$ permuting the field slots. This group acts by $\sigma(\partial_i) = \partial_{\sigma(i)}$, $\sigma \in S_N$. Each slot carries a holomorphic and an anti-holomorphic index together so $S_N$ acts diagonally on the combined space: $\partial_i\bar\partial_j \to \partial_{\sigma(i)}\bar\partial_{\sigma(j)}$. It is therefore the joint monomial that must be symmetric and not each chiral half on its own. This restricts us to the invariant subspace $\left[ \mathrm{Sym}^m(W_N) \otimes \mathrm{Sym}^{\bar{m}}(\bar{W}_N) \right]^{S_N}$.

The second condition is that a genuine primary must not be a global descendant. A state built from the total derivative $L_{-1}^{\rm tot} = \sum_i \partial_i \equiv \partial_{\rm tot}$ (or $\bar L_{-1}^{\rm tot}=\bar\partial_{\rm tot}$) is by definition a descendant, and such states must be removed from the space of states when constructing primaries. One way to do this is to restrict to the subspace orthogonal to $\partial_{\rm tot}$ (and $\bar\partial_{\rm tot}$). But this is not enough: a state can also be a linear combination of a primary and a descendant, without itself being obtained by acting with $L_{-1}^{\rm tot}$ on some other state, so restricting to the orthogonal subspace does not automatically single out the primaries. The reason is that the orthogonality within the coordinate space $W_N$ is a purely algebraic notion, and need not coincide with the orthogonality under the actual CFT inner product. The correct and general statement is that a state is a true primary, or highest-weight state, only if it is annihilated by the total raising operators,
\begin{equation}
L_1^{\text{tot}} |\psi\rangle = 0, \qquad \bar{L}_1^{\text{tot}} |\psi\rangle  = 0~,
\label{eq:annihilation_of_psi}
\end{equation}
where $L_1^{\rm tot}=\sum_i L_1^{(i)}$ acts on each field slot through the usual global commutation relations. Hence, the physical primary space is strictly the simultaneous kernel of both raising operators inside the $S_N$-invariant subspace,
\begin{equation}
\mathcal{P}^{(N)}_{(m,\bar{m})} = \ker(L_1^{\text{tot}}) \cap \ker(\bar{L}_1^{\text{tot}}) \Big|_{\left[ \mathrm{Sym}^m(W_N) \otimes \mathrm{Sym}^{\bar{m}}(\bar{W}_N) \right]^{S_N}}~.
\end{equation}
Note that the condition \eqref{eq:annihilation_of_psi} only demands invariance under $L_1^{\rm tot}$ and $\bar L_1^{\rm tot}$: this defines a \emph{global} primary, in the sense of the global $SL(2,\mathbb{R})_L\times SL(2,\mathbb{R})_R$ subalgebra generated by $\{L_{-1},L_0,L_1\}$, and not a primary of the full Virasoro algebra, which would additionally require $L_n^{\rm tot}|\psi\rangle=0$ for every $n\geq 2$. 

In general, the physical highest-weight states are given by specific $S_N$-invariant linear combinations of derivative monomials evaluated at a coincident point:
and explicit primary operators take the form
\begin{equation}
[\mathcal{O}^N]_{m,\bar{m}}(z, \bar{z}) = \sum_{\vec{k}, \bar{\vec{k}}} C_{\vec{k}, \bar{\vec{k}}} \left[ \partial_1^{k_1}\bar{\partial}_1^{\bar{k}_1}\mathcal{O} \dots \partial_N^{k_N}\bar{\partial}_N^{\bar{k}_N}\mathcal{O} \right]
\end{equation}
with the coefficients $C_{\vec k,\bar{\vec k}}$ fixed by imposing \eqref{eq:annihilation_of_psi}. Because $L_1^{\rm tot}$ acts on a stack of derivatives with a coefficient that depends on both the derivative order and the weight $h$, these coefficients are generally nontrivial functions of $h$, and finding them means solving \eqref{eq:annihilation_of_psi} directly rather than reading them off some simpler substitution. We carry this out explicitly in the next appendix for some low-lying cases.

In this appendix, by contrast, our goal is only to \emph{count} the states, i.e.\ to compute $d^{(N)}_{m,\bar m}=\dim\mathcal{P}^{(N)}_{(m,\bar m)}$, and for this purpose there is a considerable shortcut. A convenient way to remove the total-derivative direction from $W_N$ is to write $W_N = V_N \oplus \mathbb{C}\,\partial_{\rm tot}$, where $V_N$ is the $(N-1)$-dimensional subspace satisfying the zero-sum condition $\sum_{i=1}^N \partial_i = 0$ (and similarly $\bar W_N = \bar V_N \oplus \mathbb{C}\,\bar\partial_{\rm tot}$). This is a purely linear, $h$-independent condition on the coefficients, and by itself it says nothing about whether $L_1^{\rm tot}$ annihilates a state built from $V_N$. So while $V_N$ need not coincide with $\ker(L_1^{\rm tot})$ as a subspace, it always has the same dimension, and that is all we need in this appendix: we apply the zero-sum constraint to build representative operators that capture the correct primary counts. The explicit construction of genuine multi-trace primaries is carried out in the next appendix.

We now use this to compute $d^{(N)}_{m,\bar m}$ via the character of the $S_N$ action on $\mathrm{Sym}^m(V_N)\otimes\mathrm{Sym}^{\bar m}(\bar V_N)$ and also explicitly count the states by imposing both bosonic symmetry and zero sum constraint.

Let $P^{(N)}$ be the projector onto the $S_N$-invariant part of this space, given by the group average \cite{Hamermesh}
\begin{equation}
P^{(N)} = \frac{1}{|S_N|} \sum_{\sigma \in S_N} \rho(\sigma)~,
\end{equation}
where $\rho(\sigma)$ is the representation matrix of $\sigma$ on $\mathrm{Sym}^m(V_N)\otimes\mathrm{Sym}^{\bar m}(\bar V_N)$. Since the dimension of a subspace equals the trace of the projector onto it,
\begin{equation}
d_{m, \bar{m}}^{(N)} = \mathrm{Tr}(P^{(N)}) = \frac{1}{|S_N|} \sum_{\sigma \in S_N} \mathrm{Tr}\left(\rho(\sigma)\right)~.
\end{equation}
Because $S_N$ acts diagonally, $\rho(\sigma) = \rho_m(\sigma)\otimes\bar\rho_{\bar m}(\sigma)$, and the trace of a tensor product factorizes into a product of traces on each chiral factor:
\begin{equation}
\mathrm{Tr}\left(\rho(\sigma)\right) = \mathrm{Tr}_{\mathrm{Sym}^m(V_N)}\left(\rho_m(\sigma)\right) \times \mathrm{Tr}_{\mathrm{Sym}^{\bar{m}}(\bar{V}_N)}\left(\bar{\rho}_{\bar{m}}(\sigma)\right)~.
\end{equation}
Writing $\chi_m(\sigma) = \mathrm{Tr}_{\mathrm{Sym}^m(V_N)}\left(\rho_m(\sigma)\right)$ and $\chi_{\bar{m}}(\sigma) = \mathrm{Tr}_{\mathrm{Sym}^{\bar{m}}(\bar{V}_N)}\left(\bar{\rho}_{\bar{m}}(\sigma)\right)$ for the two chiral characters, the counting formula becomes
\begin{equation}
d_{m, \bar{m}}^{(N)} = \frac{1}{|S_N|} \sum_{\sigma \in S_N} \chi_m(\sigma) \chi_{\bar{m}}(\sigma)~.
\label{count}
\end{equation}

Below we compute $\chi_m(\sigma)$ explicitly for the reduced representation $\mathrm{Sym}^m(V_N)$ and work out $d^{(N)}_{m,\bar m}$ for $N=2$ and $N=3$. This confirms the correct multiplicities at each level: the actual highest-weight vectors realizing these multiplicities are then constructed explicitly for a few values of $m$ and $\bar m$, by solving \eqref{eq:annihilation_of_psi}, in the next appendix.

\subsection{$N=2$}

The action of the holomorphic partial derivative $\partial_i$ on the operator $\mathcal{O}$ is represented by the variable $x_i$, and similarly, the action of the anti-holomorphic derivative $\bar{\partial}_i$ is represented by $y_i$. For $N=2$, the single-particle derivative vector spaces $W$ and $\bar{W}$ are spanned by $\{x_1, x_2\}$ and $\{y_1, y_2\}$, respectively. To remove global translation descendants and count the primaries, one can typically restricts the analysis to the tensor product of the zero-sum subspaces, $V \otimes \bar{V}$, defined by:
\begin{equation}
V = \{(x_1, x_2) \in \mathbb{C}^2 \mid x_1 + x_2 = 0\}, \quad \bar{V} = \{(y_1, y_2) \in \mathbb{C}^2 \mid y_1 + y_2 = 0\}.
\end{equation}
However, to make the $S_2$ action on the monomial space manifest, we will instead first construct the unconstrained monomial space $\text{Sym}^m(W) \otimes \text{Sym}^{\bar m}(\bar{W})$ and subsequently impose both the $S_2$ invariance and the zero-sum constraints. 

The symmetric group $S_2$ acts on these derivative coordinates by permuting their indices, such that $\sigma \cdot x_i = x_{\sigma(i)}$ for $\sigma \in S_2$. Because $S_2$ is generated entirely by the single transposition $(12)$, testing and ensuring invariance under this generator guarantees invariance under the entire group.

To compute the number of physical primary states at derivative level $(m, \bar{m})$ for $N=2$, we evaluate the invariant subspace dimension by working directly with the unconstrained parent vector spaces, $W_2$ and $\bar{W}_2$, and implementing both the symmetric group action and the zero-sum constraints explicitly.

The full parent spaces are spanned by the independent single-particle coordinates:
\begin{equation}
W_2 = \mathrm{span}\{x_1, x_2\}, \quad \bar{W}_2 = \mathrm{span}\{y_1, y_2\}
\end{equation}
The symmetric group $S_2 = \{e, (12)\}$ acts on these spaces by permuting the particle indices ($x_1 \leftrightarrow x_2$ and $y_1 \leftrightarrow y_2$). 
At derivative level $(m, \bar{m})$, a general unconstrained configuration belongs to the tensor product of the symmetric powers, $\mathrm{Sym}^m(W_2) \otimes \mathrm{Sym}^{\bar{m}}(\bar{W}_2)$. A basis for this unconstrained parent space consists of the monomial products $x_1^a x_2^{m-a} y_1^b y_2^{\bar{m}-b}$. 
To isolate the primary configurations, we enforce the zero-sum constraints directly on the coordinates:
\begin{equation}
x_2 = -x_1, \quad y_2 = -y_1
\end{equation}
Substituting these constraints into the general monomial configurations projects the entire space down to a 1-dimensional physical subspace spanned by the single independent basis element:
\begin{equation}
\psi_{(m, \bar{m})} = x_1^m y_1^{\bar{m}}
\end{equation}

We determine the characters $\chi_{m,\bar{m}}(\sigma)$ of this physical subspace by tracking how the group elements transform this basis element under the zero-sum constraint:
The identity transformation leaves all coordinates unchanged:
\begin{equation}
e \cdot \left( x_1^m y_1^{\bar{m}} \right) = x_1^m y_1^{\bar{m}} = 1 \cdot \left( x_1^m y_1^{\bar{m}} \right) \implies \chi_{m,\bar{m}}(e) = 1,
\end{equation}
The transposition swaps the particle labels ($x_1 \leftrightarrow x_2$ and $y_1 \leftrightarrow y_2$). Applying the group action first yields:
\begin{equation}
(12) \cdot \left( x_1^m y_1^{\bar{m}} \right) = x_2^m y_2^{\bar{m}}
\end{equation}
Next, pulling this transformed state back onto the physical subspace by enforcing the constraints $x_2 = -x_1$ and $y_2 = -y_1$ gives:
\begin{equation}
(-x_1)^m (-y_1)^{\bar{m}} = (-1)^m (-1)^{\bar{m}} x_1^m y_1^{\bar{m}} = (-1)^{m+\bar{m}} \left( x_1^m y_1^{\bar{m}} \right) \implies \chi_{m,\bar{m}}\big((12)\big) = (-1)^{m+\bar{m}},
\end{equation}
To find the dimension of the $S_2$-invariant subspace, we average these characters over the group:
\begin{equation}
\begin{split}
d^{(2)}_{m, \bar{m}} &= \frac{1}{|S_2|} \sum_{\sigma \in S_2} \chi_{m, \bar{m}}(\sigma) \\
&= \frac{1}{2} \left[ \chi_{m, \bar{m}}(e) + \chi_{m, \bar{m}}\big((12)\big) \right] \\
&= \frac{1 + (-1)^{m + \bar{m}}}{2}
\end{split}
\end{equation}
This evaluation recovers the selection rule:
\begin{equation}
d^{(2)}_{m, \bar{m}} = \begin{cases} 
1 & \text{if } m + \bar{m} \text{ is even} \\ 
0 & \text{if } m + \bar{m} \text{ is odd} 
\end{cases}
\end{equation}

Now, we analyze the combined space $\mathrm{Sym}^m(V) \otimes \mathrm{Sym}^{\bar{m}}(\bar{V})$ for each pair $(m, \bar{m})$ within the range $0 \le m, \bar{m} \le 1$ to count the primary states. Due to the structural symmetry under swapping the holomorphic and anti-holomorphic sectors, the pairs $(m, \bar{m})$ and $(\bar{m}, m)$ form isomorphic representations under the action of $S_2$ and are evaluated together.

\subsubsection*{Case $(m, \bar{m}) = (0, 0)$}

For the base case $(m, \bar{m}) = (0, 0)$, the unconstrained parent space $\mathrm{Sym}^0(W) \otimes \mathrm{Sym}^0(\bar{W})$ is spanned by the constant scalar $1$. A general vector in this space is simply:
\begin{equation}
v = c \cdot 1
\label{eq:N2-00}
\end{equation}
Checking for $S_2$ invariance, the transposition $(12)$ acts trivially on scalars, meaning $(12) \cdot v = v \implies c = c$.

Next, we impose the zero-sum constraints $x_2 = -x_1$ and $y_2 = -y_1$ to project onto the subspace $[\mathrm{Sym}^0(V) \otimes \mathrm{Sym}^0(\bar{V})]^{S_2}$. Because there are no coordinate variables present at this derivative level, the constraint has no effect. The dimension of this physical space remains exactly $1$. This perfectly matches the selection rule $d^{(2)}_{0,0} = 1$. Setting $c=1$, the translation into the derivative framework yields the bare multi-trace operator:
\begin{equation}
:\mathcal{O}(z_1, \bar{z}_1)\mathcal{O}(z_2, \bar{z}_2):
\end{equation}

\subsubsection*{Cases $(1, 0)$ and $(0, 1)$}

We will demonstrate the $(1, 0)$ case; the $(0, 1)$ case follows identically using the anti-holomorphic sector. The unconstrained parent space $\mathrm{Sym}^1(W) \otimes \mathrm{Sym}^0(\bar{W})$ is spanned by the single-particle coordinates $\{x_1, x_2\}$. A general vector in this space is written as:
\begin{equation}
v = c_1 x_1 + c_2 x_2
\end{equation}
To ensure $S_2$ invariance, we apply the group generator $(12)$, which swaps the particle indices $x_1 \leftrightarrow x_2$:
\begin{equation}
(12) \cdot v = c_1 x_2 + c_2 x_1
\end{equation}
Equating this transformed vector back to $v$ requires $c_1 = c_2 \equiv c$. The invariant vector in the parent space is thus restricted to:
\begin{equation}
v_{\text{inv}} = c(x_1 + x_2)
\label{eq:N2-10-inv}
\end{equation}
To find the dimension of the physical subspace $[\mathrm{Sym}^1(V) \otimes \mathrm{Sym}^0(\bar{V})]^{S_2}$, we apply the zero-sum constraint $x_2 = -x_1$ to this invariant configuration:
\begin{equation}
v_{\text{phys}} = c(x_1 - x_1) = 0
\end{equation}
The invariant state vanishes entirely for any choice of $c$. The dimension of the physical space is $0$. This matches the selection rule $d^{(2)}_{1,0} = 0$. Consequently, there is no primary operator at this derivative level.

\subsubsection*{Case $(1, 1)$}
For the case $(1, 1)$, the unconstrained parent space $\mathrm{Sym}^1(W) \otimes \mathrm{Sym}^1(\bar{W})$ is spanned by the tensor products of the single-particle coordinates: $\{x_1 \otimes y_1, x_1 \otimes y_2, x_2 \otimes y_1, x_2 \otimes y_2\}$, we will eventually drop the tensor product symbol $\otimes$ for simplicity. A general vector in this space is:
\begin{equation}
v = c_{11} x_1 y_1 + c_{12} x_1 y_2 + c_{21} x_2 y_1 + c_{22} x_2 y_2
\end{equation}
We test for invariance under the diagonal action of the generator $(12)$, which simultaneously swaps $x_1 \leftrightarrow x_2$ and $y_1 \leftrightarrow y_2$:
\begin{equation}
(12) \cdot v = c_{11} x_2 y_2 + c_{12} x_2 y_1 + c_{21} x_1 y_2 + c_{22} x_1 y_1
\end{equation}
Equating this to the original vector $v$ and matching the linearly independent monomials yields the coefficient constraints $c_{11} = c_{22} \equiv \alpha$ and $c_{12} = c_{21} \equiv \beta$. The general invariant vector in the parent space is therefore:
\begin{equation}
v_{\text{inv}} = \alpha(x_1 y_1 + x_2 y_2) + \beta(x_1 y_2 + x_2 y_1)
\label{eq:N2-11-inv}
\end{equation}
To determine the dimension of the physical subspace $[\mathrm{Sym}^1(V) \otimes \mathrm{Sym}^1(\bar{V})]^{S_2}$, we impose the zero-sum constraints $x_2 = -x_1$ and $y_2 = -y_1$:
\begin{align*}
v_{\text{phys}} &= \alpha\big(x_1 y_1 + (-x_1)(-y_1)\big) + \beta\big(x_1(-y_1) + (-x_1)y_1\big) \\
&= \alpha(2 x_1 y_1) + \beta(-2 x_1 y_1) \\
&= 2(\alpha - \beta)x_1 y_1
\end{align*}
Because the physical state depends entirely on the single parameter combination $(\alpha - \beta)$, the dimension of the space is $1$. This successfully matches the selection rule $d^{(2)}_{1,1} = 1$.

To write down the corresponding representative operator for the primary, we return to our invariant parent vector $v_{\text{inv}}$ and pick coefficient values such that $\alpha \neq \beta$  (e.g., setting $\alpha = 1$ and $\beta = -1$):
\begin{equation}
v_{\text{inv}} = (x_1 y_1 + x_2 y_2) - (x_1 y_2 + x_2 y_1)
\end{equation}
Mapping these coordinate polynomials back to their corresponding slot derivatives yields the operator:
\begin{equation}
(\partial_1 \bar{\partial}_1 + \partial_2 \bar{\partial}_2 - \partial_1 \bar{\partial}_2 - \partial_2 \bar{\partial}_1):\mathcal{O}(z_1, \bar{z}_1)\mathcal{O}(z_2, \bar{z}_2):
\end{equation}

\subsection{$N=3$}

For $N=3$, the parent single-particle derivative vector spaces are $W_3 = \mathrm{span}\{x_1, x_2, x_3\}$ and $\bar{W}_3 = \mathrm{span}\{y_1, y_2, y_3\}$. The symmetric group $S_3$ acts on these coordinates by permuting indices. To count the physical multiplicities, we evaluate the dimension of the $S_3$-invariant subspace of the quotient space defined by the zero-sum constraints $x_3 = -x_1 - x_2$ and $y_3 = -y_1 - y_2$.

Following the procedure developed for $N=2$, the dimension of the physical primary space is given by the average of the character products over the conjugacy classes of $S_3$. A basis for the $m$-th symmetric power of the physical subspace $\mathrm{Sym}^m(V)$ consists of the $m+1$ monomials $\phi_a = x_1^a x_2^{m-a}$, with $a \in \{0, \dots, m\}$. 

While the single-sector characters $\chi_m(\sigma)$ of $\mathrm{Sym}^m(V)$ can be determined by explicitly computing the transformation matrices $\rho_m(\sigma)$ and their traces for each conjugacy class under the constraint $x_3 = -x_1 - x_2$, this computation becomes unwieldy as $N$ increases. Instead, a more general approach bypasses explicit coordinate manipulations by focusing directly on the eigenvalues of the group elements acting on the $2$-dimensional physical subspace $V$. The natural $3$-dimensional permutation representation space $\mathbb{C}^3$ decomposes into a direct sum of irreducible representations under $S_3$:
\begin{equation}
\mathbb{C}^3 = \mathbf{1} \oplus V~,
\end{equation}
where $\mathbf{1}$ is the $1$-dimensional trivial representation spanned by the uniform translation vector $(1, 1, 1)^T$, and $V$ is the $2$-dimensional physical zero-sum plane. If a group element $\sigma$ acts on the $2$-dimensional subspace $V$ with eigenvalues $\lambda_1$ and $\lambda_2$, its action on the basis elements of the $m$-th symmetric power $\mathrm{Sym}^m(V)$ yields the eigenvalues $\lambda_1^k \lambda_2^{m-k}$ for $k = 0, 1, \dots, m$. Because the character is the trace of the transformation matrix, $\chi_m(\sigma)$ is simply the sum of these eigenvalues:
\begin{equation}
\chi_m(\sigma) = \sum_{k=0}^m \lambda_1^k \lambda_2^{m-k}~.
\end{equation}
When $\lambda_1 \neq \lambda_2$, this expression evaluates as a closed-form geometric series:
\begin{equation}
\chi_m(\sigma) = \frac{\lambda_1^{m+1} - \lambda_2^{m+1}}{\lambda_1 - \lambda_2}~.
\end{equation}

We find the specific eigenvalues for each conjugacy class by analyzing their action on $\mathbb{C}^3$ and projecting out the trivial translation component along $(1,1,1)^T$.

\textbf{The identity $e$:} The identity transformation leaves all components invariant. On the full $3$-dimensional space, its eigenvalues are $\{1, 1, 1\}$. Removing the eigenvalue $\lambda = 1$ associated with the invariant translation vector leaves the eigenvalues on the physical subspace $V$ as:
\begin{equation}
\lambda_1 = 1, \quad \lambda_2 = 1~.
\end{equation}
Substituting these into the symmetric power trace sum yields:
\begin{equation}
\chi_m(e) = \sum_{k=0}^m (1)^k (1)^{m-k} = m + 1~.
\end{equation}

\textbf{The transpositions $(12)$:} The transposition $(12)$ swaps the first two particle labels. Because it satisfies $\sigma^2 = e$, its eigenvalues on $\mathbb{C}^3$ must be roots of unity satisfying $\lambda^2 = 1$, which gives $\lambda = \pm 1$. The trace of $(12)$ on $\mathbb{C}^3$ is $1$ because it leaves the third coordinate $x_3$ unchanged. Since the trivial translation vector $(1, 1, 1)^T$ is symmetric under the swap, it carries the eigenvalue $+1$. To match the total $3$-dimensional trace of $1$, the remaining eigenvalues on the $2$-dimensional subspace $V$ must be:
\begin{equation}
\lambda_1 = 1, \quad \lambda_2 = -1~.
\end{equation}
Evaluating the symmetric power character sum with these eigenvalues gives an alternating series:
\begin{equation}
\chi_m\big((12)\big) = \sum_{k=0}^m (1)^k (-1)^{m-k} = \frac{1 + (-1)^m}{2}~,
\end{equation}
which vanishes for odd $m$ and equals $1$ for even $m$.

\textbf{The $3$-cycles $(123)$:} The 3-cycle $(123)$ permutes the coordinates cyclically. Because it satisfies $\sigma^3 = e$, its eigenvalues on $\mathbb{C}^3$ are the three complex cube roots of unity: $\{1, \omega, \omega^2\}$, where $\omega = e^{i2\pi/3}$. The uniform translation vector $(1, 1, 1)^T$ is invariant under cyclic shifts and accounts for the eigenvalue $1$. Consequently, the eigenvalues restricted to the physical zero-sum subspace $V$ are the non-trivial roots:
\begin{equation}
\lambda_1 = \omega = e^{i\frac{2\pi}{3}}, \quad \lambda_2 = \omega^2 = e^{-i\frac{2\pi}{3}}~.
\end{equation}
Since $\lambda_1 \neq \lambda_2$, we evaluate the character using the geometric quotient formula:
\begin{equation}
\chi_m\big((123)\big) = \frac{\left(e^{i\frac{2\pi}{3}}\right)^{m+1} - \left(e^{-i\frac{2\pi}{3}}\right)^{m+1}}{e^{i\frac{2\pi}{3}} - e^{-i\frac{2\pi}{3}}}~.
\end{equation}
Using Euler's identity to express the complex exponentials as sines, the expression simplifies to a ratio of trigonometric functions:
\begin{equation}
\chi_m\big((123)\big) = \frac{\sin\left(\frac{2\pi(m+1)}{3}\right)}{\sin\left(\frac{2\pi}{3}\right)}~.
\end{equation}
This ratio generates a periodic sequence of length $3$ as $m$ increases:
\begin{equation}
\chi_m\big((123)\big) = \begin{cases} 
1 & \text{if } m \equiv 0 \pmod 3 \\ 
-1 & \text{if } m \equiv 1 \pmod 3 \\ 
0 & \text{if } m \equiv 2 \pmod 3 
\end{cases}~.
\end{equation}

To find the dimension $d^{(3)}_{m, \bar{m}}$ of the $S_3$-invariant subspace, we average these product characters over the group, accounting for the sizes of the conjugacy classes:
\begin{equation}
\begin{split}
d^{(3)}_{m, \bar{m}} &= \frac{1}{|S_3|} \sum_{\sigma \in S_3} \chi_m(\sigma) \chi_{\bar{m}}(\sigma) \\
&= \frac{1}{6} \left[ 1 \cdot \chi_m(e)\chi_{\bar{m}}(e) + 3 \cdot \chi_m\big((12)\big)\chi_{\bar{m}}\big((12)\big) + 2 \cdot \chi_m\big((123)\big)\chi_{\bar{m}}\big((123)\big) \right]~.
\end{split}
\end{equation}
Substituting our explicit trace derivations into the group average yields the complete formula for the number of physical primary states:
\begin{equation}
d^{(3)}_{m,\bar m} = \frac{(m+1)(\bar m+1)}{6} +\frac{1}{2}\,\mathbf{1}_{m\ {\rm even}}\,\mathbf{1}_{\bar m\ {\rm even}} +\frac{1}{3}\,c_m c_{\bar m}~,
\label{eq:appA_d3_formula}
\end{equation}
where
\begin{equation}
c_m=
\begin{cases}
1, & m\equiv 0\pmod 3~,\\
-1, & m\equiv 1\pmod 3~,\\
0, & m\equiv 2\pmod 3~.
\end{cases}
\end{equation}

We now evaluate the combined space $\mathrm{Sym}^m(V) \otimes \mathrm{Sym}^{\bar{m}}(\bar{V})$ for each pair $(m, \bar{m})$ to count the primary states.

\subsubsection*{Case (1,1)}

We begin by writing the most general vector in the unconstrained space $\text{Sym}^1(W) \otimes \text{Sym}^1(\bar{W})$. This space is spanned by 9 monomials, so the general vector has 9 arbitrary coefficients:
\begin{align}
v &= c_{11}x_1y_1 + c_{22}x_2y_2 + c_{33}x_3y_3 + c_{12}x_1y_2 + c_{21}x_2y_1 \nonumber \\
&\quad + c_{23}x_2y_3 + c_{32}x_3y_2 + c_{31}x_3y_1 + c_{13}x_1y_3
\end{align}
Applying the transposition $(12)$, which swaps $1 \leftrightarrow 2$ and leaves $3$ invariant, we get:
\begin{align}
v^{(12)} &= c_{11}x_2y_2 + c_{22}x_1y_1 + c_{33}x_3y_3 + c_{12}x_2y_1 + c_{21}x_1y_2 \nonumber \\
&\quad + c_{23}x_1y_3 + c_{32}x_3y_1 + c_{31}x_3y_2 + c_{13}x_2y_3
\end{align}
For $v$ to be an invariant state under $S_3$, it must be symmetric, so $v = v^{(12)}$. Equating the coefficients of corresponding monomials yields the first set of constraints:
\begin{equation}
c_{11} = c_{22}, \quad c_{12} = c_{21}, \quad c_{23} = c_{13}, \quad c_{32} = c_{31}
\end{equation}
Applying the 3-cycle $(123)$, which maps $1 \to 2 \to 3 \to 1$, gives:
\begin{align}
v^{(123)} &= c_{11}x_2y_2 + c_{22}x_3y_3 + c_{33}x_1y_1 + c_{12}x_2y_3 + c_{21}x_3y_2 \nonumber \\
&\quad + c_{23}x_3y_1 + c_{32}x_1y_3 + c_{31}x_1y_2 + c_{13}x_2y_1
\end{align}
Equating $v = v^{(123)}$ and feeding in the relations from $v=v^{(12)}$ fully locks the coefficients into exactly two independent classes:
\begin{align}
c_{11} &= c_{22} = c_{33} \equiv c_A \\
c_{12} &= c_{21} = c_{23} = c_{13} = c_{31} = c_{32} \equiv c_B
\end{align}
The general symmetric vector is now strictly a linear combination of two invariant orbits:
\begin{equation}
v_{\text{inv}} = c_A (x_1y_1 + x_2y_2 + x_3y_3) + c_B (x_1y_2 + x_2y_1 + x_2y_3 + x_3y_2 + x_3y_1 + x_1y_3)
\label{eq:N3-11-inv-vector}
\end{equation}
Let,
\begin{equation}
\begin{aligned}
O_1 &= x_1y_1 + x_2y_2 + x_3y_3 \\
O_2 &= x_1y_2 + x_2y_1 + x_2y_3 + x_3y_2 + x_3y_1 + x_1y_3
\end{aligned}
\label{eq:N3-11-orbits}
\end{equation}
We substitute $x_3 = -x_1 - x_2$ and $y_3 = -y_1 - y_2$ into $v_{\text{inv}}$, this gives $O_1 \to 2x_1y_1 + 2x_2y_2 + x_1y_2 + x_2y_1$ and $O_2 \to -2x_1y_1 - 2x_2y_2 - x_1y_2 - x_2y_1$.
Therefore, $O_2 = -O_1$. Substituting this back into the general vector gives:
\begin{equation}
v_{\text{phys}} = (c_A - c_B) O_1
\end{equation}
Since $c_A$ and $c_B$ are arbitrary, $(c_A - c_B)$ acts as a single overall normalization constant. There is 1 independent {representative of the} physical {primary} state. Translating $O_1$ to derivatives yields the state:
\begin{equation}
 (\partial_1 \bar{\partial}_1 + \partial_2 \bar{\partial}_2 + \partial_3 \bar{\partial}_3) : \mathcal{O}(z_1, \bar{z}_1)\mathcal{O}(z_2, \bar{z}_2)\mathcal{O}(z_3, \bar{z}_3) :
\end{equation}

\subsubsection*{Case (2,2)}

The general vector in the 36-dimensional unconstrained space $\text{Sym}^2(W) \otimes \text{Sym}^2(\bar{W})$ has 36 arbitrary coefficients, written as:
\begin{equation}
v = \sum_{i \leq j, \, k \leq l} c_{ij,kl} (x_i x_j)(y_k y_l)
\end{equation}
Enforcing $v = v^{(12)}$ requires symmetry between indices 1 and 2 (e.g., $c_{11,11} = c_{22,22}$ and $c_{11,33} = c_{22,33}$). Enforcing $v = v^{(123)}$ requires cyclic equality across all three indices. Applying both conditions partitions the 36 unconstrained coefficients into exactly 8 independent equivalence classes. Setting the coefficients within each class to a single variable $c_A \dots c_H$, the general symmetric vector simplifies directly into the sum of the 8 orbits:
\begin{align}
v_{\text{inv}} &= c_A O_A + c_B O_B + c_C O_C + c_D O_D + c_E O_E + c_F O_F + c_G O_G + c_H O_H
\label{eq:N3-22-inv-vector}
\end{align}
where the orbits are:
\begin{equation}
\begin{aligned}
O_A &= x_1^2 y_1^2 + x_2^2 y_2^2 + x_3^2 y_3^2 \\
O_B &= x_1^2 y_2^2 + x_1^2 y_3^2 + x_2^2 y_1^2 + x_2^2 y_3^2 + x_3^2 y_1^2 + x_3^2 y_2^2 \\
O_C &= x_1^2 y_1 y_2 + x_1^2 y_3 y_1 + x_2^2 y_1 y_2 + x_2^2 y_2 y_3 + x_3^2 y_2 y_3 + x_3^2 y_3 y_1 \\
O_D &= x_1^2 y_2 y_3 + x_2^2 y_3 y_1 + x_3^2 y_1 y_2 \\
O_E &= x_1 x_2 y_3^2 + x_2 x_3 y_1^2 + x_3 x_1 y_2^2 \\
O_F &= x_1 x_2 y_1^2 + x_1 x_2 y_2^2 + x_2 x_3 y_2^2 + x_2 x_3 y_3^2 + x_3 x_1 y_3^2 + x_3 x_1 y_1^2 \\
O_G &= x_1 x_2 y_1 y_2 + x_2 x_3 y_2 y_3 + x_3 x_1 y_3 y_1 \\
O_H &= x_1 x_2 y_1 y_3 + x_1 x_2 y_2 y_3 + x_2 x_3 y_2 y_1 + x_2 x_3 y_3 y_1 + x_3 x_1 y_3 y_2 + x_3 x_1 y_1 y_2
\end{aligned}
\label{eq:N3-22-inv-orbits}
\end{equation}
We project onto the physical subspace by substituting $x_3 = -x_1 - x_2$ and $y_3 = -y_1 - y_2$. By expanding the polynomials into the independent basis $\{x_1^2y_1^2, x_2^2y_2^2, \dots\}$, we find the algebraic identities linking the orbits:
\begin{equation}
O_C = -O_A, \quad O_F = -O_A, \quad O_D = O_E = -O_H = \frac{1}{2}(O_A - O_B), \quad O_G = \frac{3}{4}O_A - \frac{1}{4}O_B
\end{equation}
Substituting these identities back into the general vector $v_{\text{inv}}$ gives:
\begin{align}
v_{\text{phys}} &= c_A O_A + c_B O_B + c_C(-O_A) + c_D \left[\frac{1}{2}(O_A - O_B)\right] + c_E \left[\frac{1}{2}(O_A - O_B)\right] \nonumber \\
&\quad + c_F(-O_A) + c_G\left[\frac{3}{4}O_A - \frac{1}{4}O_B\right] + c_H\left[-\frac{1}{2}(O_A - O_B)\right]
\end{align}
Grouping the terms by the independent orbits $O_A$ and $O_B$:
\begin{align}
v_{\text{phys}} &= \left( c_A - c_C + \frac{1}{2}c_D + \frac{1}{2}c_E - c_F + \frac{3}{4}c_G - \frac{1}{2}c_H \right) O_A \nonumber \\
&\quad + \left( c_B - \frac{1}{2}c_D - \frac{1}{2}c_E - \frac{1}{4}c_G + \frac{1}{2}c_H \right) O_B 
\end{align}

Since the original $c_i$ coefficients are arbitrary, the terms in the large parentheses simply define two new independent, arbitrary constants, let's call them $K_1$ and $K_2$. The physical subspace is exactly 2-dimensional:
\begin{equation}
v_{\text{phys}} = K_1 O_A + K_2 O_B
\end{equation}

The two linearly independent states correspond to the following operators built directly from $O_A$ and $O_B$ acing on the three copies of the field $\mathcal{O}$:
\begin{align}
&\partial_1^2 \bar{\partial}_1^2 + \partial_2^2 \bar{\partial}_2^2 + \partial_3^2 \bar{\partial}_3^2 \\
&\partial_1^2 \bar{\partial}_2^2 + \partial_1^2 \bar{\partial}_3^2 + \partial_2^2 \bar{\partial}_1^2 + \partial_2^2 \bar{\partial}_3^2 + \partial_3^2 \bar{\partial}_1^2 + \partial_3^2 \bar{\partial}_2^2
\end{align}

\section{Explicit Construction of Multi-Trace Primaries}
\label{sec:honest_primaries}

In Appendix~\ref{sec:Multi_trace_primary_construction} we constructed, for $N=2$ and $N=3$, the $S_N$-invariant subspace at a few selected derivative levels $(m,\bar m)$ \sout{each derivative level $(m,\bar m)$}, and used the zero-sum quotient to compute its dimension $d^{(N)}_{m,\bar m}$. As emphasized there, this dimension count is rigorous, but the explicit vector singled out by the zero-sum substitution need not be annihilated by $L_1^{\rm tot}$ and $\bar L_1^{\rm tot}$, and therefore need not be the actual primary. In this appendix we return to the $S_N$-invariant vectors already identified in Appendix~\ref{sec:Multi_trace_primary_construction} and impose the honest primary condition $L_1^{\rm tot}|\psi\rangle=\bar L_1^{\rm tot}|\psi\rangle=0$ directly, using the operator algebra rather than any coordinate substitution.

The key tool throughout this appendix is the exact action of the total raising operator $L_1^{\rm tot}$ on a state built from several powers of $L_{-1}$ on a single field slot. We derive this here explicitly, since it is this identity that determines whether a given $S_N$-invariant state is an honest primary.

Consider a single field slot with $p$ holomorphic derivatives acting on it, $L_{-1}^p|\mathcal O\rangle$, where $\mathcal O$ has weight $h$. Since the state $|\mathcal O\rangle$ is itself primary, $L_1|\mathcal O\rangle=0$, so acting with $L_1$ on the descendant reduces entirely to a nested commutator:
\begin{equation}
L_1 L_{-1}^p |\mathcal O\rangle = [L_1, L_{-1}^p]\,|\mathcal O\rangle~.
\end{equation}
Using the standard operator identity $[A,B^p] = \sum_{k=0}^{p-1} B^k [A,B] B^{p-1-k}$ with $A=L_1$, $B=L_{-1}$, and the $\mathfrak{sl}(2)$ commutation relation $[L_1,L_{-1}]=2L_0$, this becomes
\begin{equation}
L_1 L_{-1}^p |\mathcal O\rangle = \sum_{k=0}^{p-1} L_{-1}^k [L_1,L_{-1}] L_{-1}^{p-1-k}|\mathcal O\rangle = 2\sum_{k=0}^{p-1} L_{-1}^k L_0 L_{-1}^{p-1-k}|\mathcal O\rangle~.
\end{equation}
Each term now has $L_0$ sandwiched between two stacks of lowering operators. Using $L_0 L_{-1}^{n}|\mathcal O\rangle = (h+n)L_{-1}^{n}|\mathcal O\rangle$ (which follows from $[L_0,L_{-1}]=L_{-1}$ applied $n$ times) with $n=p-1-k$, every term in the sum reduces to the same descendant $L_{-1}^{p-1}|\mathcal O\rangle$, with a $k$-dependent coefficient:
\begin{equation}
L_1 L_{-1}^p |\mathcal O\rangle = 2\sum_{k=0}^{p-1} \big(h+p-1-k\big)\,L_{-1}^{p-1}|\mathcal O\rangle~.
\end{equation}
The remaining sum is a simple arithmetic series over $k=0,\ldots,p-1$:
\begin{equation}
\sum_{k=0}^{p-1}\big(h+p-1-k\big) = p(h+p-1) - \sum_{k=0}^{p-1}k = p(h+p-1) - \frac{p(p-1)}{2}~.
\end{equation}
Collecting terms,
\begin{equation}
2\left[p(h+p-1) - \frac{p(p-1)}{2}\right] = p\big(2h+p-1\big)~,
\end{equation}
so that
\begin{equation}
L_1\,(L_{-1})^{p}\,|\mathcal{O}\rangle \;=\; p\,(2h+p-1)\,(L_{-1})^{p-1}\,|\mathcal{O}\rangle~,
\label{eq:raising-identity}
\end{equation}
and identically for the anti-holomorphic generators\footnote{Note that we are working with scalar seed primaries, so $h= {\bar h}$.}, $\bar L_1(\bar L_{-1})^{p}|\mathcal O\rangle = p(2h+p-1)(\bar L_{-1})^{p-1}|\mathcal O\rangle$. Translating back to field-derivative notation, acting with $L_1^{(i)}$ on a single field slot carrying $p$ derivatives gives
\begin{equation}
L_1^{(i)} \left( \partial_i^p \mathcal{O}(z_i, \bar{z}_i) \right) = p(2h+p-1)\, \partial_i^{p-1} \mathcal{O}(z_i, \bar{z}_i)~.
\end{equation}

At $p=1$ the coefficient is simply $2h$ -- the same number regardless of which slot it came from. For $p\geq 2$, the coefficient $p(2h+p-1)$ depends nontrivially on both $h$ and the derivative order $p$, and it is precisely this $h$-dependence that the flat, coordinate-level zero-sum substitution of Appendix~\ref{sec:Multi_trace_primary_construction} never captured. Equation~\eqref{eq:raising-identity} is the only input we need below: at each level, we act with $L_1^{\rm tot}=\sum_i L_1^{(i)}$ (and $\bar L_1^{\rm tot}$) on the $S_N$-invariant vectors already constructed in Appendix~\ref{sec:Multi_trace_primary_construction}, term by term, using \eqref{eq:raising-identity} slot by slot, and solve for the coefficients that make the result vanish.
\subsection{$N=2$}

\subsubsection*{Case $(0,0)$}
The invariant space at this level is one-dimensional, spanned by Eq.~\eqref{eq:N2-00}. Since $L_1^{(i)}$ annihilates a field slot with no derivatives on it, $L_1^{\rm tot}|\psi\rangle=0$ trivially, for any $c$. This is consistent with $d^{(2)}_{0,0}=1$, and the bare product $:\mathcal O\mathcal O:$ is already the honest primary.

\subsubsection*{Case $(1,0)$ (and $(0,1)$)}
The $S_2$-invariant state of Eq.~\eqref{eq:N2-10-inv} is $|\psi\rangle_{\rm inv} = c(\partial_1+\partial_2):\mathcal O\mathcal O:$. Applying Eq.~\eqref{eq:raising-identity} at $p=1$ to each term,
\begin{equation}
L_1^{\rm tot}|\psi\rangle_{\rm inv} = c\,(2h+2h):\mathcal O\mathcal O: = 4hc:\mathcal O\mathcal O:~,
\end{equation}
which vanishes only for $c=0$ at generic $h$. This matches $d^{(2)}_{1,0}=0$: there is no primary at this level.

\subsubsection*{Case $(1,1)$}
The $S_2$-invariant state of Eq.~\eqref{eq:N2-11-inv} is
\begin{equation}
|\psi\rangle_{\rm inv} = \big[\alpha(\partial_1\bar\partial_1+\partial_2\bar\partial_2) + \beta(\partial_1\bar\partial_2+\partial_2\bar\partial_1)\big]:\mathcal O\mathcal O:~.
\end{equation}
Applying Eq.~\eqref{eq:raising-identity} at $p=1$, $L_1^{(i)}$ removes $\partial_i$ and leaves the anti-holomorphic derivative untouched, with coefficient $2h$ on every term:
\begin{equation}
L_1^{\rm tot}|\psi\rangle_{\rm inv} = 2h(\alpha+\beta)(\bar\partial_1+\bar\partial_2):\mathcal O\mathcal O:~,
\end{equation}
which vanishes iff $\alpha=-\beta$. By the symmetry of $|\psi\rangle_{\rm inv}$ under exchanging holomorphic and anti-holomorphic indices, $\bar L_1^{\rm tot}|\psi\rangle_{\rm inv}=0$ gives the same condition. This fixes a one-dimensional solution space, matching $d^{(2)}_{1,1}=1$. Choosing $\alpha=1,\beta=-1$, the honest primary is
\begin{equation}
(\partial_1\bar\partial_1+\partial_2\bar\partial_2-\partial_1\bar\partial_2-\partial_2\bar\partial_1):\mathcal O(z_1,\bar z_1)\mathcal O(z_2,\bar z_2):~.
\end{equation}
This coincides with the state obtained from the zero-sum substitution earlier. This is not a coincidence: at $p=1$ the coefficient in Eq.~\eqref{eq:raising-identity} is the constant $2h$, independent of which slot it acts on, so at this level the honest primary condition and the flat zero-sum condition happen to select the same subspace. As we see below, this ceases to hold already at $N=3$.

\subsection{$N=3$}

\subsubsection*{Case $(1,1)$}
The two $S_3$-invariant orbits of Eq.~\eqref{eq:N3-11-inv-vector}, translated to operator language, are
\begin{equation}
O_1 = \sum_i L_{-1}^{(i)}\bar L_{-1}^{(i)}|\mathcal O\mathcal O\mathcal O\rangle~,\qquad
O_2 = \sum_{i\neq j} L_{-1}^{(i)}\bar L_{-1}^{(j)}|\mathcal O\mathcal O\mathcal O\rangle~,
\end{equation}
with $|\psi\rangle_{\rm inv}=c_AO_1+c_BO_2$. There, the zero-sum substitution gave $O_2=-O_1$, so the primary it selects is always proportional to $O_1$ alone, regardless of $c_A,c_B$. We now check this directly against Eq.~\eqref{eq:raising-identity}. Since $L_1^{(i)}$ only ever acts on the holomorphic $L_{-1}^{(i)}$ sitting on its own slot,
\begin{equation}
L_1^{\rm tot}\,O_1 = \sum_i 2h\,\bar L_{-1}^{(i)}|\mathcal O\mathcal O\mathcal O\rangle = 2h\,\bar L_{-1}^{\rm tot}|\mathcal O\mathcal O\mathcal O\rangle~,
\end{equation}
while for $O_2$, each fixed $j$ receives a contribution from the two choices of $i\neq j$,
\begin{equation}
L_1^{\rm tot}\,O_2 = \sum_{i\neq j} 2h\,\bar L_{-1}^{(j)}|\mathcal O\mathcal O\mathcal O\rangle = 4h\,\bar L_{-1}^{\rm tot}|\mathcal O\mathcal O\mathcal O\rangle~.
\end{equation}
Hence
\begin{equation}
L_1^{\rm tot}\big(c_AO_1+c_BO_2\big) = 2h(c_A+2c_B)\,\bar L_{-1}^{\rm tot}|\mathcal O\mathcal O\mathcal O\rangle~,
\end{equation}
which vanishes iff $c_A=-2c_B$. By the symmetry of $O_1,O_2$ under exchanging chiralities, the same relation gives $\bar L_1^{\rm tot}|\psi\rangle_{\rm inv}=0$. This confirms $d^{(3)}_{1,1}=1$ while showing explicitly that the zero-sum primary is not the true one. Choosing $c_B=1,c_A=-2$, the honest primary is
\begin{equation}
:\Big[(\partial_1\bar\partial_2+\partial_1\bar\partial_3+\partial_2\bar\partial_1+\partial_2\bar\partial_3+\partial_3\bar\partial_1+\partial_3\bar\partial_2) - 2(\partial_1\bar\partial_1+\partial_2\bar\partial_2+\partial_3\bar\partial_3)\Big]\mathcal O(z_1,\bar z_1)\mathcal O(z_2,\bar z_2)\mathcal O(z_3,\bar z_3):~.
\end{equation}

\subsubsection*{Case $(2,2)$}
Eq.~\eqref{eq:N3-22-inv-orbits} gave the eight $S_3$-invariant orbits $O_A,\ldots,O_H$ spanning the invariant space $|\psi\rangle_{\rm inv}=\sum_X c_X O_X$ at this level. We now impose Eq.~\eqref{eq:raising-identity} directly on this eight-dimensional space, without using the algebraic identities among $O_A,\ldots,O_H$ that followed there from the zero-sum substitution.

Write $\alpha\equiv 2h$ and $\beta\equiv 2(2h+1)$ for the $p=1$ and $p=2$ coefficients in Eq.~\eqref{eq:raising-identity}. Since $L_1^{(i)}$ only acts on the holomorphic derivatives on slot $i$, each orbit maps under $L_1^{\rm tot}$ to a combination of the four level-$(1,2)$ invariant sums
\begin{equation}
Q_1 = \sum_i \partial_i\bar\partial_i^2~,\quad
Q_2 = \sum_{i\neq j}\partial_i\bar\partial_j^2~,\quad
Q_3 = \sum_{i\neq j}\partial_i\bar\partial_i\bar\partial_j~,\quad
Q_4 = \partial_1\bar\partial_2\bar\partial_3+\partial_2\bar\partial_1\bar\partial_3+\partial_3\bar\partial_1\bar\partial_2~.
\end{equation}
Orbits built from a doubled derivative $x_i^2$ (i.e.\ $O_A,O_B,O_C,O_D$) always yield the $p=2$ coefficient $\beta$; orbits built from $x_ix_j$ (i.e.\ $O_E,O_F,O_G,O_H$) always yield the $p=1$ coefficient $\alpha$, with a factor of $2$ whenever two slots can contribute. Carrying this out slot by slot gives
\begin{align}
L_1^{\rm tot}O_A &= \beta\,Q_1\,, & L_1^{\rm tot}O_E &= \alpha\,Q_2\,,\notag\\
L_1^{\rm tot}O_B &= \beta\,Q_2\,, & L_1^{\rm tot}O_F &= \alpha\,(2Q_1+Q_2)\,,\notag\\
L_1^{\rm tot}O_C &= \beta\,Q_3\,, & L_1^{\rm tot}O_G &= \alpha\,Q_3\,,\notag\\
L_1^{\rm tot}O_D &= \beta\,Q_4\,, & L_1^{\rm tot}O_H &= \alpha\,(Q_3+2Q_4)\,.
\end{align}
Since $Q_1,\ldots,Q_4$ are linearly independent, $L_1^{\rm tot}|\psi\rangle_{\rm inv}=0$ splits into four conditions:
\begin{align}
\beta\,c_A + 2\alpha\,c_F &= 0\,, \label{eq:22-c1}\\
\beta\,c_B + \alpha\,c_E + \alpha\,c_F &= 0\,, \label{eq:22-c2}\\
\beta\,c_C + \alpha\,c_G + \alpha\,c_H &= 0\,, \label{eq:22-c3}\\
\beta\,c_D + 2\alpha\,c_H &= 0\,. \label{eq:22-c4}
\end{align}
Swapping $x_i\leftrightarrow y_i$ throughout exchanges $L_1^{\rm tot}\leftrightarrow\bar L_1^{\rm tot}$ and acts on the orbits as $O_A,O_B,O_G,O_H\to$ themselves and $O_C\leftrightarrow O_F$, $O_D\leftrightarrow O_E$. Applying this swap to Eqs.~\eqref{eq:22-c1}--\eqref{eq:22-c4} gives the four conditions from $\bar L_1^{\rm tot}|\psi\rangle_{\rm inv}=0$:
\begin{align}
\beta\,c_A + 2\alpha\,c_C &= 0\,, \label{eq:22-cb1}\\
\beta\,c_B + \alpha\,c_C + \alpha\,c_D &= 0\,, \label{eq:22-cb2}\\
\beta\,c_F + \alpha\,c_G + \alpha\,c_H &= 0\,, \label{eq:22-cb3}\\
\beta\,c_E + 2\alpha\,c_H &= 0\,. \label{eq:22-cb4}
\end{align}
Comparing Eq.~\eqref{eq:22-c1} with Eq.~\eqref{eq:22-cb1} forces $c_C=c_F$; comparing Eq.~\eqref{eq:22-c4} with Eq.~\eqref{eq:22-cb4} forces $c_D=c_E$. With these, Eqs.~\eqref{eq:22-c2}--\eqref{eq:22-cb2} and \eqref{eq:22-c3}--\eqref{eq:22-cb3} coincide, leaving four independent equations. Solving with $c_A,c_H$ free:
\begin{equation}
c_C=c_F=-\frac{\beta}{2\alpha}c_A\,,\qquad
c_D=c_E=-\frac{2\alpha}{\beta}c_H\,,\qquad
c_G=\frac{\beta^2}{2\alpha^2}c_A-c_H\,,\qquad
c_B=\frac{1}{2}c_A+\frac{2\alpha^2}{\beta^2}c_H\,.
\end{equation}
This gives a two-dimensional solution space, matching $d^{(3)}_{2,2}=2$. Substituting $\alpha=2h,\beta=2(2h+1)$ and taking $(c_A,c_H)=(1,0)$ and $(0,1)$ (cleared of denominators) gives the two honest primaries at level $(2,2)$:
\begin{align}
\Psi_1 &= 2h^2\,O_A + h^2\,O_B - h(2h+1)\,(O_C+O_F) + (2h+1)^2\,O_G\,,\\
\Psi_2 &= 2h^2\,O_B - 2h(2h+1)\,(O_D+O_E) - (2h+1)^2\,O_G + (2h+1)^2\,O_H\,,
\end{align}
where it is understood that these expressions act on three copies of $\mathcal{O}$ at a coincident point. Both $\Psi_1$ and $\Psi_2$ are annihilated by $L_1^{\rm tot}$ and $\bar L_1^{\rm tot}$ for any $h$, by construction. Neither is proportional to $O_A$ or $O_B$ alone, setting all other coefficients other than $c_A$ and $c_B$ to zero in Eqs.~\eqref{eq:22-c1}--\eqref{eq:22-c4} forces $c_A=0$ or $c_B=0$ respectively, since $\beta\neq 0$ for generic $h$. This shows explicitly that the states $O_A=\sum_i\partial_i^2\bar\partial_i^2$ and $O_B=\sum_{i\neq j}\partial_i^2\bar\partial_j^2$ identified by the zero-sum method are not honest primaries: they carry nonzero images under $L_1^{\rm tot}$, proportional to $Q_1$ and $Q_2$ respectively, and cancelling these residuals requires mixing in the six additional orbits $O_C,\ldots,O_H$ that the zero-sum substitution discarded.

\section{Fixed-$N$ Partition Functions: Windings vs. Young Diagrams}
\label{sec:schur_weyl_review}

In this Appendix we will organize the multi-winding Wilson loops using the representation theory of the permutation group $S_N$. This is not strictly necessary to understand the multi-trace structure, but it has the utility that: (a) it allows us to write down explicit expressions for $\chi_N$ that are difficult to obtain by direct Plethystic exponentiation at high $N$, and (b) it gives us some intuition on how the left and chiral channels combine.

\subsection{Winding Basis and the Cycle-Index Formula}

Let $W$ be the single-particle state space, and let $U$ be a diagonal operator on $W$ with eigenvalues $\{x_i\}_{i \in I}$. A single particle winding $n$ times is written as the power sum:
\begin{equation}
    p_n(x) := \sum_{i \in I} x_i^n = \mathrm{Tr}_W\, U^n .
    \label{eq:powersums}
\end{equation}
This may seem a bit abstract, so for our purposes, it is good to keep the following example in mind: $W$ is the single-particle Hilbert space $\mathcal{H}_1$ and $U = e^{-\beta H}$, so that the $n=1$ power sum reproduces the single-particle partition function,
\begin{equation}
    p_1\!\left(e^{-\beta E}\right) = \sum_{i \in I} e^{-\beta E_i} = \mathrm{Tr}_{\mathcal{H}_1}\, e^{-\beta H} .
\end{equation}
Specifically, our interest will be in the case where $W$ is the global conformal module. This state space is spanned by a primary and its descendants under $L_{-1}$. The eigenvalues are explicitly given by
\begin{equation}
x_\ell := q^{h+\ell}, \ \ 
\ell \ge0,
\label{eq:xy_def}
\end{equation}
where $h$ denotes the conformal weight of the primary, and $q = e^{2 \pi i \tau}$. We will also need their anti-chiral counterparts: $y_{\bar\ell} := \bar q^{h+\bar\ell}$ with $\bar \ell \ge0$. $W$ can be understood as a single \emph{chiral} Hilbert space, $W=\mathcal H_L$ or $W=\mathcal H_R$, or as the full one-particle space $\mathcal H_1=\mathcal H_L\otimes\mathcal H_R$, depending on context. Correspondingly, $U$ denotes the chiral operator acting on $\mathcal H_L$ or $\mathcal H_R$, generating the eigenvalues $q^{h+\ell}$ or $\bar q^{h+\bar\ell}$ introduced above. In the next subsection we will keep track of the detailed chiral-antichiral decomposition.

To find the trace over the symmetric (bosonic) subspace $\mathrm{Sym}^N(W) \subset W^{\otimes N}$, we project down from the full tensor-product space using the standard symmetrization operator. Let $|\phi\rangle$ be any vector in $W^{\otimes N}$, and let $P$ be a linear operator from $W^{\otimes N}$ to $W^{\otimes N}$. In order for $P$ to project $|\phi\rangle$ to $\mathrm{Sym}^N(W)$, $P$ must satisfy the following two conditions:
\begin{align}
    \widehat{\sigma}\, P\, |\phi\rangle &= P\, |\phi\rangle \qquad \text{for every } \sigma \in S_N, \label{eq:invariance_condition}\\
    P\, |\phi\rangle &= |\phi\rangle \qquad \text{if } |\phi\rangle \in \mathrm{Sym}^N(W).
    \label{eq:fixing_condition}
\end{align}
where $\widehat{\sigma}$ stands for the explicit implementation of the abstract permutation $\sigma$ on the $N$-particle Hilbert space. The standard operator satisfying these conditions is the group average
\begin{equation}
    P = \frac{1}{|S_N|} \sum_{\sigma \in S_N} \widehat{\sigma}.
    \label{eq:projector_def}
\end{equation}
To see that Eq.~\eqref{eq:invariance_condition} holds, take any $\sigma' \in S_N$ and act on $P$:
\begin{equation}
    \widehat{\sigma}'\, P = \frac{1}{|S_N|} \sum_{\sigma \in S_N} \widehat{\sigma}'\,\widehat{\sigma} = \frac{1}{|S_N|} \sum_{\sigma'' \in S_N} \widehat{\sigma}'' = P,
\end{equation}
where we used that $\sigma'' = \sigma'\sigma$ is itself an element of $S_N$, and (because $\sigma \mapsto \sigma'\sigma$ is a bijection of $S_N$ onto itself) summing over all $\sigma \in S_N$ is equivalent to summing over all $\sigma'' \in S_N$. So the relabeled sum is identical to the original, giving $\widehat{\sigma}'P = P$ for every $\sigma' \in S_N$.
To see that Eq.~\eqref{eq:fixing_condition} holds, suppose $|\phi\rangle$ is already symmetric, i.e.\ $\widehat{\sigma}|\phi\rangle = |\phi\rangle$ for all $\sigma \in S_N$. Then
\begin{equation}
    P\,|\phi\rangle = \frac{1}{N!} \sum_{\sigma \in S_N} \widehat{\sigma}\,|\phi\rangle = \frac{1}{N!} \sum_{\sigma \in S_N} |\phi\rangle = \frac{1}{N!}\cdot N!\, |\phi\rangle = |\phi\rangle.
\end{equation}
Since $P$ satisfies both defining conditions, it is precisely the projector from $W^{\otimes N}$ onto $\mathrm{Sym}^N(W)$. It follows that the trace over the symmetric subspace can be computed as a trace over the full space $W^{\otimes N}$ weighted by $P$:
\begin{equation}
    \mathrm{Tr}_{\mathrm{Sym}^N(W)}\, U^{\otimes N}
    = \frac{1}{N!} \sum_{\sigma \in S_N} \mathrm{Tr}_{W^{\otimes N}}\!\left(\widehat{\sigma}\cdot U^{\otimes N}\right),
    \label{eq:projector_step}
\end{equation}
where $U^{\otimes N}$ denotes $N$ copies of $U$, each acting diagonally on one factor of the $N$-particle Hilbert space $W^{\otimes N}$. From here on, we will drop the tensor-power notation and simply write $U$ in place of $U^{\otimes N}$, with the number of copies understood from context (i.e.\ matching the number of particles $N$ under consideration).

It is useful to make the following aside for clarity in later discussions. The analogous projector onto the antisymmetric (fermionic) subspace $\wedge^N(W)$ is obtained by weighting the sum by the sign of the permutation,
\begin{equation}
    P_{\text{antisym}} = \frac{1}{N!}\sum_{\sigma\in S_N}\mathrm{sgn}(\sigma)\,\widehat\sigma,
    \label{eq:projector_fermion}
\end{equation}
where $\mathrm{sgn}(\sigma)$ is the sign of $\sigma$, equal to $+1$ if $\sigma$ is a product of an even number of transpositions and $-1$ if odd. This gives
\begin{equation}
    \mathrm{Tr}_{\wedge^N(W)}\, U
    = \frac{1}{N!} \sum_{\sigma \in S_N} \mathrm{sgn}(\sigma)\,\mathrm{Tr}_{W^{\otimes N}}\!\left(\widehat{\sigma}\cdot U\right).
    \label{eq:projector_step_fermion}
\end{equation}

When we trace over a single permutation $\sigma$, we only get non-zero results if the states stay exactly the same under that permutation.\footnote{For example, consider a system of $6$ particles and a permutation $\sigma = (1\,2\,3)(4\,5)(6) \in S_6$ acting on a state $|i_1,i_2,i_3,i_4,i_5,i_6\rangle$:
\begin{equation}
    \widehat{\sigma}\,|i_1,i_2,i_3,i_4,i_5,i_6\rangle = |i_2,i_3,i_1,i_5,i_4,i_6\rangle.
\end{equation}
The inner product $\langle i_1,i_2,i_3,i_4,i_5,i_6 \,|\, i_2,i_3,i_1,i_5,i_4,i_6\rangle$ is then non-zero only if $i_1=i_2=i_3$ and $i_4=i_5$.} This means all particles in a specific permutation cycle must be in the exact same state. Physically, a cycle of length $|c|$ connects these identical particles into one loop that winds the thermal circle $|c|$ times. Because independent cycles act on particles in different states, the trace splits into a product:
\begin{equation}
\mathrm{Tr}_{W^{\otimes N}} (\widehat{\sigma} \cdot U) = \prod_{c\in \mathrm{cycles}(\sigma)} \mathrm{Tr}_W\,U^{|c|}.
\label{eq:trace_split}
\end{equation}

The trace formula in Eq.~\eqref{eq:trace_split} depends only on the \emph{cycle type} of $\sigma$ (i.e.\ the multiset of cycle lengths). Two permutations with different cycle structure on different particles, but the same cycle lengths, give identical contributions to the trace. For example, take $\sigma = (1\,2\,3)(4\,5)(6) \in S_6$ and $\sigma' = (1\,2\,4)(3\,6)(5) \in S_6$. Both have cycle type $[3,2,1]$ (one 3-cycle, one 2-cycle, one fixed point), even though they act on entirely different subsets of particles: $\sigma$ groups particles $\{1,2,3\}$, $\{4,5\}$, $\{6\}$, while $\sigma'$ groups $\{1,2,4\}$, $\{3,6\}$, $\{5\}$. Despite this, both give
\begin{equation}
    \mathrm{Tr}_{W^{\otimes 6}}(\widehat{\sigma}\cdot U^{\otimes 6}) = \mathrm{Tr}_{W^{\otimes 6}}(\widehat{\sigma'}\cdot U^{\otimes 6}) = p_3(x)\,p_2(x)\,p_1(x),
\end{equation}
see \eqref{eq:powersums} for definition of $p_i(x)$. Since Eq.~\eqref{eq:trace_split} only sees the cycle lengths $\{3,2,1\}$, common to both, this lets us simplify \eqref{eq:projector_step} by grouping permutations with the same cycle structure into conjugacy classes. We track this structure using a partition $\mu$ of $N$ (denoted $\mu \vdash N$). A partition $\mu$ is sometimes written as $(1^{m_1}2^{m_2}...)$ or as $\prod_{k\ge1} k^{m_k}$, where $m_k$ is the number of cycles of length $k$. The permutation $\sigma = (1\,2\,3)(4\,5)(6)$ has cycle type $\mu = (1^1\, 2^1 \, 3^1)$.

The number of permutations in a class $\mu$ is $N!$ divided by a redundancy factor $z_\mu$:
\begin{equation}
|C_\mu| = \frac{N!}{z_\mu}, \qquad \text{where} \quad z_\mu := \prod_{k\ge1} k^{m_k}\,m_k!.
\label{eq:zmu_def}
\end{equation}
The $z_\mu$ factor corrects for the overcounting coming from cyclic permutations within a cycle of length $k$, which gives a factor of $k$. If there are $m_k$ such cycles, this becomes $k^{m_k}$, the permutation of these $m_k$ cycles among themselves is also a redundancy, which gives the factor $m_k!$. Changing the sum from all $N!$ permutations to just a sum over the partitions $\mu$ gives
\begin{equation}
    \sum_{\sigma \in S_N} \mathrm{Tr}_{W^{\otimes N}}\!\left(\widehat{\sigma}\cdot U^{\otimes N}\right) = \sum_{\mu \vdash N} |C_\mu|\, \prod_{k\ge1} \left(\mathrm{Tr}_W\, U^k\right)^{m_k},
\end{equation}
using \eqref{eq:projector_step}, this gives the cycle-index formula:
\begin{equation}
\mathrm{Tr}_{\mathrm{Sym}^N(W)} U
=
\sum_{\mu\vdash N}\frac{1}{z_\mu}\prod_{k\ge1}\big(\mathrm{Tr}_W\,U^k\big)^{m_k}.
\label{eq:cycle_index_general}
\end{equation}

In our physical setup, $W = \mathcal{H}_1$ and $U = q^{L_0}\bar{q}^{\bar{L}_0}$, so $\mathrm{Tr}_W U^k = \chi_1(q^k,\bar q^k)$. This gives the general multi-particle character for a fixed $N$:
\begin{equation}
\chi_N(q,\bar q)
=
\sum_{\mu\vdash N}\frac{1}{z_\mu}\prod_{k\ge1}\chi_1(q^k,\bar q^k)^{m_k}.
\label{eq:chiN_cycle_general}
\end{equation}
For $N=2$ and $N=3$, this evaluates to:
\begin{equation}
\chi_2(q,\bar q)
=
\frac{1}{2}\Big(\chi_1(q,\bar q)^2+\chi_1(q^2,\bar q^2)\Big),
\label{eq:chi2_cycle}
\end{equation}
\begin{equation}
\chi_3(q,\bar q)
=
\frac{1}{6}\Big(\chi_1(q,\bar q)^3
+3\,\chi_1(q,\bar q)\chi_1(q^2,\bar q^2)
+2\,\chi_1(q^3,\bar q^3)\Big).
\label{eq:chi3_cycle}
\end{equation}
Physically, a term like $\chi_1(q, \bar{q})^2$ means single winding for two particles in same state. A term like $\chi_1(q^2, \bar{q}^2)$ means one particle winding twice. The cycle-index formula simply adds these loops together with the correct statistical weights $z_\mu$.

\subsection{Young-Diagram Basis}

Both the Wilson-spool picture and the Wilson tower picture organize the partition function as a sum over conformal characters: the Wilson spool writes it as a sum over multi-winding characters, while the Wilson tower assigns a single-winding character to each multi-trace primary. There is a third, equivalent way to organize the same data, in terms of the irreducible representations (irreps) of $S_N$.

In the previous section we took the Hilbert space of $N$ particles and projected onto its symmetric subspace to obtain the bosonic partition function. Here we consider a refinement of that setup, in which the single-particle Hilbert space is itself a tensor product of left- and right-moving chiral pieces, $\mathcal H_1=\mathcal H_L\otimes\mathcal H_R$. In this chiral setting, projecting $\mathcal H_L$ and $\mathcal H_R$ onto their symmetric subspaces separately is not enough to build the bosonic space. Instead, the full $N$-particle state is bosonic only if $\mathcal H_L$ and $\mathcal H_R$ transform in the \emph{same} irreducible representation of $S_N$. To make this precise, we first need a projector onto an arbitrary irrep $\lambda$, generalizing the symmetric and antisymmetric projectors already introduced.

Just as summing $\sigma\in S_N$ with weight $+1$ gives the symmetric projector in Eq.~\eqref{eq:projector_def} and weighting by $\mathrm{sgn}(\sigma)$ gives the antisymmetric projector in Eq.~\eqref{eq:projector_fermion}, weighting instead by the character\footnote{The character is defined as $\chi^\lambda(\sigma)=\operatorname{Tr}\big(D^\lambda(\sigma)\big)$, where $D^\lambda(\sigma)$ is the matrix representing the permutation $\sigma$ in the irreducible representation $\lambda$ of $S_N$.} $\chi^\lambda(\sigma)$ projects onto the isotypic component\footnote{The irrep $\lambda$ may appear multiple times in $W^{\otimes N}$, the collection is referred to as the isotypic component.} 
of $W^{\otimes N}$ transforming in $\lambda$\cite{Hamermesh}:
\begin{equation}
P_\lambda = \frac{\dim\lambda}{N!}\sum_{\sigma\in S_N}\chi^\lambda(\sigma)\,\widehat\sigma.
\end{equation}
The trace of $U^{\otimes N}$ restricted to this subspace, computed the same way as in Eq.~\eqref{eq:projector_step}, defines the Schur function $s_\lambda(x)$:
\begin{equation}
s_\lambda(x)
=
\frac{1}{\dim\lambda}\,\mathrm{Tr}_{W^{\otimes N}}\!\left(P_\lambda\,U^{\otimes N}\right)
=
\sum_{\mu\vdash N}
\frac{\chi^\lambda(\mu)}{z_\mu}\,\prod_{k\ge1}p_k(x)^{m_k},
\label{eq:frobenius_map}
\end{equation}
where $\chi^\lambda(\mu)$ is the character of $S_N$ in irrep $\lambda$ evaluated on the conjugacy class of cycle type $\mu$ (which are known numbers for $S_N$), and $p_k(x)=\mathrm{Tr}_W U^k$.

A bosonic state does not require its left- and right-moving sectors to be individually symmetric. Rather, they must transform in the same irreducible representation of $S_N$, so that the combined wavefunction is symmetric under exchange of the full particles. Using the one-particle eigenvalues from Eq.~\eqref{eq:xy_def}, the partition function over all particle numbers can be identified with a classical identity in the theory of symmetric functions. The refined Cauchy identity \cite{MacdonaldSymmetricFunctions} states that, for any two sets of variables $\{x_\ell\}$ and $\{y_{\bar\ell}\}$ and a formal parameter $t$, 
\begin{equation}
\prod_{\ell,\bar\ell\ge0}
\frac{1}{1-t\,x_\ell y_{\bar\ell}}
=
\sum_{N\ge0}
t^N
\sum_{\lambda\vdash N}
s_\lambda(x)\,
s_\lambda(y).
\label{eq:cauchy_refined}
\end{equation}
We will take the $\{x_\ell\}$ to be the chiral eigenvalues introduced in \eqref{eq:xy_def}, and $\{y_{\bar\ell}\}$ to be their anti-chiral counterparts. Comparing the right-hand side to Eqs.~\eqref{eq:Zt_def} and~\eqref{eq:chiN_def}, the coefficient of $t^N$ is exactly $\chi_N(q,\bar q)$.

Extracting the fixed-$N$ coefficient gives the character-level statement that the bosonic Hilbert space decomposes into sectors of matching left- and right-moving $S_N$ representations:
\begin{equation}
\chi_N(q,\bar q)
=
\sum_{\lambda\vdash N}
s_\lambda(x)\,
s_\lambda(y).
\label{eq:chiN_schur}
\end{equation}
For $N=2,3$ this reads
\begin{align}
\chi_2
&=
s_{(2)}(x)s_{(2)}(y)
+
s_{(1,1)}(x)s_{(1,1)}(y),
\nonumber\\[2pt]
\chi_3
&=
s_{(3)}(x)s_{(3)}(y)
+
s_{(2,1)}(x)s_{(2,1)}(y)
+
s_{(1,1,1)}(x)s_{(1,1,1)}(y),
\label{eq:chi23_young}
\end{align}
with the Schur functions written in terms of power sums via Eq.~\eqref{eq:frobenius_map} as
\begin{equation}
s_{(2)}
=
\frac{p_1^2+p_2}{2},
\qquad
s_{(1,1)}
=
\frac{p_1^2-p_2}{2},
\label{eq:schur_N2}
\end{equation}
\begin{equation}
s_{(3)}
=
\frac{p_1^3+3p_1p_2+2p_3}{6},
\qquad
s_{(2,1)}
=
\frac{p_1^3-p_3}{3},
\qquad
s_{(1,1,1)}
=
\frac{p_1^3-3p_1p_2+2p_3}{6}.
\label{eq:schur_N3}
\end{equation}
Explicitly plugging these expressions into \eqref{eq:chiN_schur}, one sees that only the terms of the form $p_n(x)p_n(y)$ (which can be replaced by $\chi_1(q^n,\bar q^n)$) emerge. This reproduces the cycle-index expressions of Eqs.~\eqref{eq:chi2_cycle} and~\eqref{eq:chi3_cycle}, confirming that the Young-diagram expansion and the cycle decomposition are two equivalent descriptions of the same spectrum.

As a concrete illustration, $S_2$ has only two irreps: the symmetric $\lambda=(2)$ and the antisymmetric $\lambda=(1,1)$. The primary generating function~\eqref{eq:PN_def} then splits into two matched exchange-symmetry channels,
\begin{equation}
P_2(q,\bar q)
=
(q\bar q)^{2h}
\left[
\widehat P_{(2)}(q)\,
\widehat P_{(2)}(\bar q)
+
\widehat P_{(1,1)}(q)\,
\widehat P_{(1,1)}(\bar q)
\right],
\label{eq:P2_young_channels}
\end{equation}
where
\begin{equation}
\widehat P_{(2)}(q)
:=
(1-q)\,
q^{-2h}\,
s_{(2)}(x)
=
\frac{1}{1-q^2},
\qquad
\widehat P_{(1,1)}(q)
:=
(1-q)\,
q^{-2h}\,
s_{(1,1)}(x)
=
\frac{q}{1-q^2}.
\label{eq:P2_channel_functions}
\end{equation}
The symmetric channel generates only even values of $m$, the antisymmetric channel only odd values. Since the bosonic projection keeps only matched $S_2$ representations for the left- and right-moving sectors, the allowed states satisfy that $m$ and $\bar m$ share the same parity -- reproducing the selection rule of Eq.~\eqref{eq:d2_rule}.

\end{document}